\documentclass[prl,twocolumn,showpacs,amsmath,amssymb]{revtex4-1}
\usepackage{graphicx}
\usepackage{amsfonts}
\usepackage{amsmath}
\usepackage{amssymb}
\usepackage{color}
\usepackage{bm}
\usepackage{bbm}
\usepackage{enumerate}
\usepackage{color}
\usepackage{amsfonts, amsmath, amsthm, amssymb} 
\usepackage{array}
\usepackage[colorlinks,bookmarks=false,citecolor=blue,linkcolor=red,urlcolor=blue]{hyperref}

\usepackage[colorlinks,bookmarks=false,citecolor=blue,linkcolor=red,urlcolor=blue]{hyperref}
\usepackage[usenames,dvipsnames,svgnames,table]{xcolor}

\newcommand{\be}{\begin{equation}}
\newcommand{\bee}{\begin{equation*}}
\newcommand{\ee}{\end{equation}}
\newcommand{\eee}{\end{equation*}}
\newcommand{\bearre}{\begin{eqnarray*}}
\newcommand{\eearre}{\end{eqnarray*}}
\newcommand{\bearr}{\begin{eqnarray}}
\newcommand{\eearr}{\end{eqnarray}}

\begin{document}

\title{Entanglement entropy across the superfluid-insulator transition : a signature of bosonic criticality}

\author{Ir\'en\'ee Fr\'erot$^1$
 and Tommaso Roscilde$^{1,2}$
 }

\affiliation{$^1$ Univ Lyon, Ens de Lyon, Univ Claude Bernard, CNRS, Laboratoire de Physique, F-69342 Lyon, France}
\affiliation{$^2$ Institut Universitaire de France, 103 boulevard Saint-Michel, 75005 Paris, France}

\date{\today}

\begin{abstract}
 We study the entanglement entropy and entanglement spectrum of the paradigmatic Bose-Hubbard model, describing strongly correlated bosons on a lattice. The use of a controlled approximation - the slave-boson approach - allows us to study entanglement in all regimes of the model (and, most importantly, across its superfluid/Mott-insulator transition) at a minimal cost. We find that the area-law scaling of entanglement -- verified in all the phases -- exhibits a sharp singularity at the transition. The singularity is greatly enhanced when the transition is crossed at fixed, integer filling, due to a richer entanglement spectrum containing an additional gapless mode, which descends from the amplitude (Higgs) mode of the global excitation spectrum -- while this mode remains gapped at the generic (commensurate-incommensurate) transition with variable filling. Hence the entanglement properties contain a unique signature of the two different forms of bosonic criticality exhibited by the Bose-Hubbard model. 
\end{abstract}

\maketitle


\textit{Introduction}.---
Quantum entanglement \cite{Horodecki2009} 
offers an entirely new perspective on quantum many-body systems. In particular the scaling of the entanglement entropy (EE) of a subsystem with its size may provide a classification of known many-body phases which is alternative (or complementary) to that based on correlations properties and order parameters. But, most intriguingly, entanglement may help to see properties that traditional probes can hardly (or simply cannot) see. The prime example of this scenario is represented by topological entanglement entropy \cite{Hammaetal2005,LevinW2006,KitaevP2006}, standing as a unique defining feature of topological order.
More recently, several studies \cite{Metlitski-Grover,Alba2013,Luitzetal2015,Kulchytskyyetal2015} have focused on the Anderson ``tower of states" in the spectrum of finite-size systems \cite{Anderson1952} -- a precursor of the spontaneous breaking of a continuous symmetry in the thermodynamic limit; while such states are arguably very hard to observe spectroscopically, they leave a tangible trace in the scaling of the entanglement entropy, in the form of a universal additive logarithmic term. 
As the study of entanglement in many-body systems progresses, it is imaginable that further elusive properties will be captured by entanglement features. 

 \emph{Bose-Hubbard model and main results.} In this work we concentrate on the entanglement properties of a paradigmatic model of strongly correlated quantum particles, namely the Bose-Hubbard model \cite{Fisheretal1989}, whose Hamiltonian reads
 \begin{equation}
 {\cal H} = - J \sum_{\langle ij \rangle} \left(  b_i^{\dagger} b_j + {\rm h.c.} \right) + \sum_i \left [ \frac{U}{2} (b_i^+)^2 b_i^2 - \mu b_i^{\dagger} b_i \right]
 \end{equation}
 where $b_i, b_i^{\dagger}$ are bosonic operators, and the sums $\sum_{\langle ij \rangle}$ and $\sum_i$ run on the links and sites (respectively) of a $d$-dimensional hypercubic lattice. In the following we shall define our system on a $d$-dimensional hypertorus with size $V=L^{d-1}\times 2L$. This model is known to exhibit two different ground-states: the Mott insulator (MI) at integer filling, without condensation and with a gap to all excitations; and the superfluid condensate (SF), with a gapless Goldstone mode -- corresponding to transverse fluctuations of the order parameter -- and a gapped amplitude (or ``Higgs") mode -- corresponding to longitudinal fluctuations. 
  These two phases are connected by two different quantum phase transitions: a commensurate-incommensurate (CI) transition, driven by the ratio $\mu/U$ of the chemical potential to the repulsion, or by the ratio $J/U$ of hopping to repulsion at non-integer (variable) filling; and an O(2) transition driven by the ratio $J/U$ at integer (fixed) filling. The two transitions belong to different universality classes (Gaussian and $d$-dimensional XY model respectively), and moreover they are fundamentally different in terms of the structure of elementary excitations becoming soft at the transition: upon approaching the O(2) transition from the SF side, the amplitude mode becomes gapless in addition to the Goldstone mode, while this does not occur at the CI transition. 
  In this work we demonstrate that the entanglement entropy of a subsystem upon crossing the superfluid transition is extremely sensitive to the specific nature of the latter, and that this sensitivity descends primarily from the presence or absence of a gapless amplitude mode at the transition. Indeed we observe a ``bulk-boundary" correspondence between the (bulk) amplitude mode of the physical spectrum and a (boundary) gapped mode in the entanglement spectrum \cite{LiH2008} of a subsystem; the softening of both modes at the O(2) transition leads to a cusp singularity in the area law of entanglement entropy, absent instead at the CI transition.

  \emph{Slave-boson approach.} To investigate entanglement at the MI-SF transition, we rely on a controlled approximation, namely a quadratic expansion around the mean-field (MF) ground-state based on a slave (or Schwinger) boson representation of the boson operators \cite{Fresard1994,AltmanA2002,Dickerscheidetal2003,Altmanetal2003}. This approach allows one to reconstruct semi-quantitatively the ground-state and excitation spectrum throughout the phase diagram, including the critical points, and it can be systematically improved by a cluster approach \cite{Huerga2013}. In particular it captures correctly the existence of both the Goldstone and amplitude modes \cite{Huberetal2007,Pekkeretal2012}, as well as their fate across the superfluid-insulator transition. The slave boson (SB) approach starts from factorized ground-state Ansatz $|\Psi_{\rm MF}\rangle = \prod_i \left [ \sum_{n=0}^{n_{\rm max}} \frac{f^{(0)}_n}{\sqrt{n!}} (b_i^{\dagger})^n\right ] |0\rangle$, where $n_{\rm max}$ imposes a convenient truncation of the local Hilbert space. The variational minimization of $\langle \Psi_{\rm MF} | {\cal H} | \Psi_{\rm MF}\rangle$ amounts to the self-consistent minimization of the single-site Hamiltonian ${\cal H}_{\rm MF} = -zJ (\phi b^\dagger + \phi^*b) + \frac{U}{2} (b^\dagger)^2 b^2 -\mu b^\dagger b$  where $z=2d$, and the order parameter$\phi = \sum_n \sqrt{n} f_n^{(0)} f_{n-1}^{(0)*}$ is the ground-state expectation value of the field operator $b$, whose non-zero value distinguishes the superfluid from the insulator \cite{Sheshadrietal1993, Krutitsky2016}.
  
  \begin{figure}
		\includegraphics[width=1\linewidth]{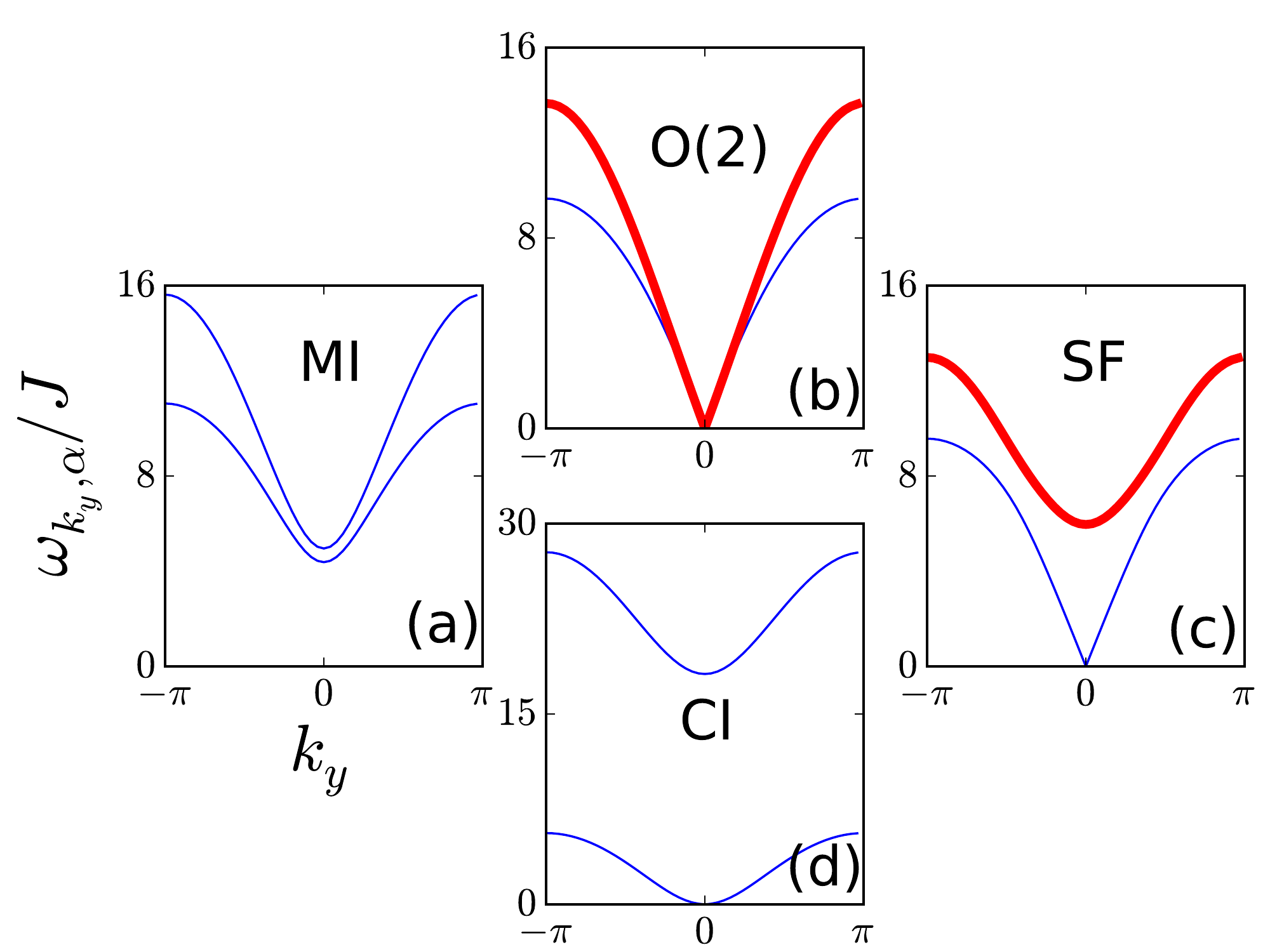}
	\includegraphics[width=1\linewidth]{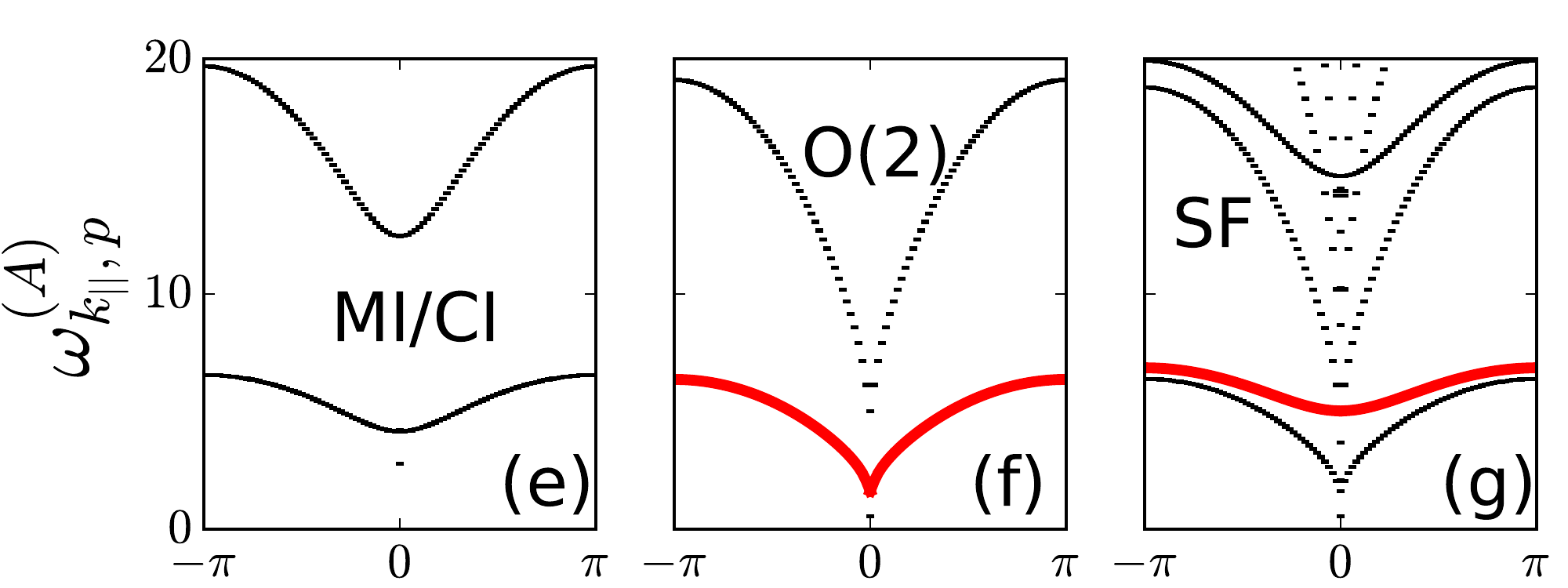}
		\caption{Physical spectrum (PS) \textit{v.s.} entanglement spectrum (ES). \textit{Upper panels}.-- PS at $k_x = 0$. Only the two lowest branches are shown. (a) $Jz/U=0.15$, $\mu/U=\sqrt 2 -1$ ; (b) $Jz/U=3-2\sqrt 2$, $\mu/U=\sqrt 2 -1$ ; (c) $Jz/U=0.2$, $\mu/U=\sqrt 2-1$ and (d) $Jz/U=0.12$, $\mu/U = 0.168$. \textit{Lower panels}.-- ES against the wave vector $k_{||}$ parallel to the  $A-B$ boundary. Only the lowest branches are shown. In the SF (MI) phase, each branch is actually two-fold (four-fold) degenerate. (e), (f), (g) correspond to the same parameters as (a), (b), (c) respectively. In all the panels, the thick red line marks the amplitude mode in the PS and in the ES.} 
 		\label{f.spectra}
              \end{figure}

  The central idea of the SB method is to enlarge the local Hilbert space, by defining $(n_{\rm max} +1)$ SB operators $\beta_{i,n}, \beta^{\dagger}_{i,n}$ with the hardcore constraint $\sum_{n} \beta^{\dagger}_{i,n} \beta_{i,n}  = 1$, such that the (physical) boson operator takes the form $b_i =  \sum_n \sqrt{n+1} ~\beta^{\dagger}_{i,n} \beta_{i,n+1}$. 
  The transformation diagonalizing the MF Hamiltonian ${\cal H}_{\rm MF}$ defines a rotation in the space of SB operators to new $\gamma$ operators, $\beta_{i,n}= \sum_{\alpha} f_n^{(\alpha)} \gamma_{i,\alpha}$ where $\alpha$ indexes the eigenstates of ${\cal H}_{\rm MF}$.  
 The SB approach consists then in ``condensing" the $\gamma_0$ bosons corresponding to the MF ground-state, $\gamma_0^{(\dagger)}\approx 1$, and treating the other bosonic fields as weak quantum fluctuations, so that one can discard terms ${\cal O}(\gamma_{\alpha\neq 0}^3)$ and higher. The resulting quadratic Hamiltonian in momentum space reads
  	${\cal H}_{2} = \frac{1}{2} \sum_{\bm k} ( {\bm{\gamma}}_{\bm k}^\dagger, {\bm{\gamma}}_{-\bm k})
			\mathcal H_{\bm k} ({\bm{\gamma}}_{\bm k} , {\bm{\gamma}}_{-\bm k}^\dagger)^{T}  $
				where ${\bm{\gamma}}_{\bm k} = V^{-1/2} \sum_i e^{-i\bm{k}\cdot \bm{r_i}} {\bm{\gamma}}_{i}$, ${\bm{\gamma}}_{i} = \{ \gamma_{\alpha\neq 0,i} \}$ is a $n_{\rm max}$-dimensional vector, and  $\mathcal H_{\bm k}$ is a $2n_{\max} \times 2n_{\max}$ matrix whose expression is given in the supplementary material \cite{SuppMat}, and which is diagonalized via a Bogoliubov transformation \cite{blaizot}.
The excitation spectrum $\omega_{\bm k,\alpha}$ exhibits $n_{\max}$ branches corresponding to the different SB flavors $\alpha$. In the insulating phase, the spectrum is gapped and the two lowest branches correspond to particle- and hole-like excitations (Fig.~\ref{f.spectra}a), while in the superfluid phase the excitation spectrum exhibits a characteristic gapless linear Goldstone mode, and a gapped amplitude mode (Fig.~\ref{f.spectra}c). At the phase transition the spectrum becomes gapless with a dispersion $\omega \sim k^z$, where $z=2$ at the CI transition (Fig.~\ref{f.spectra}d), while $z=1$ at the O(2) transition (Fig.~\ref{f.spectra}b). In the latter case the linearly dispersing mode is actually double, as the amplitude mode goes gapless at the transition. All these features, expected at the level of the field-theoretical description of the superfluid-insulator transition \cite{Fisheretal1989}, are faithfully reproduced by the SB approach. 

   
    \emph{Entanglement entropy and spectrum from slave bosons.} The main feature of the slave boson approach - namely the reduction of the Bose-Hubbard Hamiltonian to an approximate quadratic form in \emph{all} regimes -- opens the door to the calculation of the entanglement properties, as is well established in the literature \cite{PeschelE2009,BoteroR2004,Songetal2011,FrerotR2015}. In particular we shall focus on an $A$-$B$ bipartition of the $d$-dimensional hypertorus, such that the $A$-$B$ cut runs along $d-1$ dimensions, and breaks the translational invariance uniquely along the perpendicular direction. From the reduced density matrix $\rho_A$ of subsystem $A$ one can extract the von Neumann entanglement entropy and the entanglement Hamiltonian ${\cal H}_A$ such that $\rho_A = \exp(-{\cal H}_A)$. Due to Wick's theorem, such an Hamiltonian is also quadratic, ${\cal H}_A = \sum_{{\bm k}_{||},p} \omega^{(A)}_{{\bm k}_{||},p} {\lambda}_{{\bm k}_{||},p}^{\dagger} {\lambda}_{{\bm k}_{||},p}$, where $\omega^{(A)}_{{\bm k}_{||},p}$ is the (single-particle) entanglement spectrum, parametrized by the momentum ${\bm k}_{||}$ parallel to the $A$-$B$ cut, and by an additional quantum number $p$ (related both to the dynamics orthogonal to the cut, as well as to the SB flavor).    
     
     As shown in Fig.~\ref{f.spectra}(b), there is a remarkable correspondence between the  entanglement spectrum $\omega^{(A)}_{{\bm k}_{||},p}$ and  the physical spectrum $\omega_{{\bm k},\alpha}$ when crossing the superfluid-insulator transition. On the insulating side both spectra are gapped. On the superfluid side, on the other hand, the entanglement spectrum has several Goldstone-like gapless bands, vanishing as $|\ln k_{||}|^{-1}$ when $k_{||}\to 0$ \cite{Swingle2013,Metlitski-Grover,FrerotR2015}, as well as several \emph{gapped} bands. While the former are observed also within Bogoliubov or spin-wave theory \cite{FrerotR2015} - which only describe transverse fluctuations - the latter are a peculiarity of the SB approach, and they clearly correspond to ``Higgs"-like entanglement modes.  Indeed, at the O(2) transition the latter become gapless and even degenerate with the Goldstone-like modes, while they remain gapped at the CI transition. A closer inspection into the structure of the low-energy entanglement modes shows that they are spatially localized close to the $A$-$B$ boundary (see \cite{SuppMat}): hence the correspondence between the entanglement and physical spectra has the nature of a ``bulk-boundary correspondence" between the bulk excitations of the physical system, and the boundary excitations of the entanglement Hamiltonian.
     \begin{figure}
    \includegraphics[width=1\columnwidth]{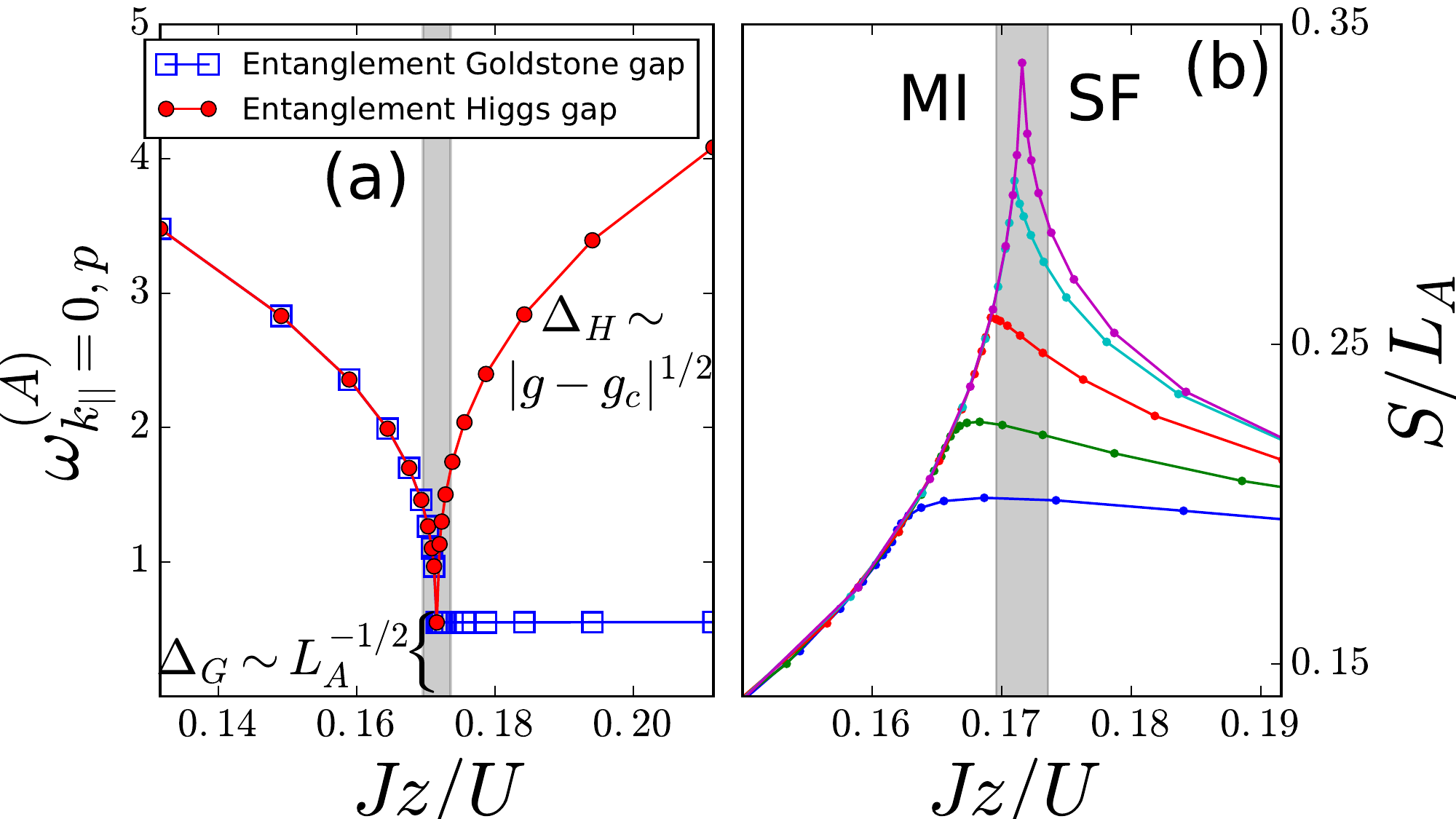}
	\includegraphics[width =1\linewidth]{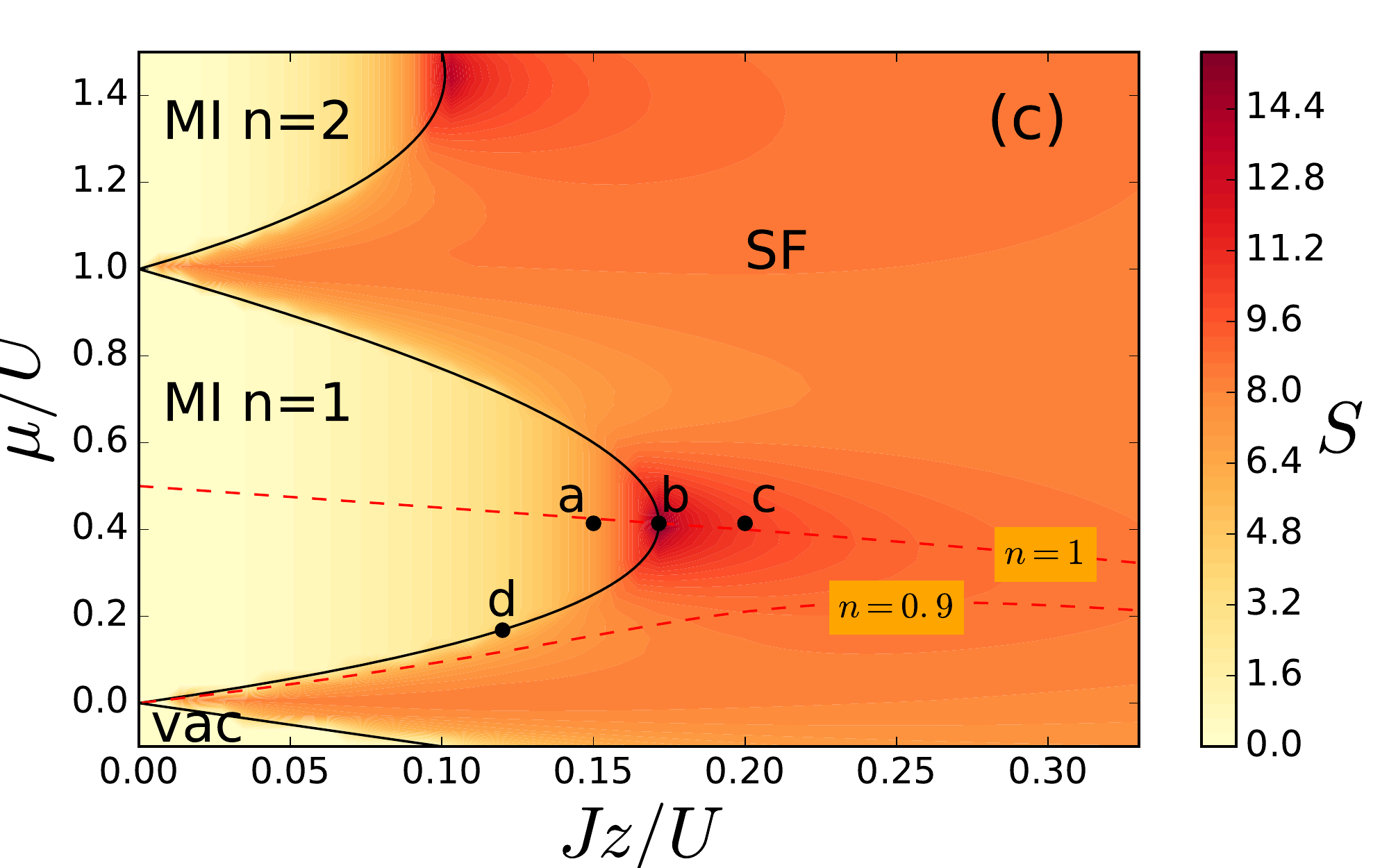}
	\caption{(a) Entanglement gaps across the SF-MI transition of the 2$d$ Bose-Hubbard model. Region $A$ corresponds to half of a $L_A\times 2L_A$ torus ($L_A=100$). The shaded area marks the region in which  $\Delta_{\rm H} \lesssim 1$. In the SF phase, the entanglement spectrum exhibits a finite-size gap $\Delta_G \sim L_A^{-1/2}$. (b) EE across the SF-MI phase transition for $\mu/U=0.3$, $0.329$, $0.357$, $0.386$ and $\sqrt{2}-1$ from bottom to top. Shaded area as in (a), and $L_A=50$ (c) EE across the phase diagram of the 2$d$ Bose-Hubbard model ($L_A=50$). a, b, c and d mark the points where the spectra are evaluated in Fig.~\ref{f.spectra}. Dashed lines correspond to constant density $n=1$ and $n=0.9$.}
	\label{f.EE}
      \end{figure}
     The study of the entanglement spectrum  allows one to regard the entanglement entropy as the thermodynamic entropy at a fictitious temperature $T=1$ for a model which develops (or does not develop) a soft mode when crossing the superfluid-insulator transition at the O(2) (or CI) point. Whenever the gapless Higgs entanglement mode develops, one may expect a singular increase in the number of low-energy states which are thermally populated. Such a singularity is indeed exhibited in Fig.~\ref{f.EE}(a), showing that the entanglement entropy develops a cusp at the O(2) transition. 
     A singular contribution from the amplitude mode is to be expected whenever the Higgs entanglement gap $\Delta_{\rm H} \lesssim 1$, such that an entanglement temperature of $T=1$ is sufficient to thermally populate the mode -- this is indeed observed in Fig.~\ref{f.EE}(b). Nonetheless, as shown in Fig.~\ref{f.EE}(c) the O(2) points at the tip of the various Mott lobes are by far the most entangled states across the whole phase diagram.  

   	\begin{figure}
	\includegraphics[width=1\columnwidth]{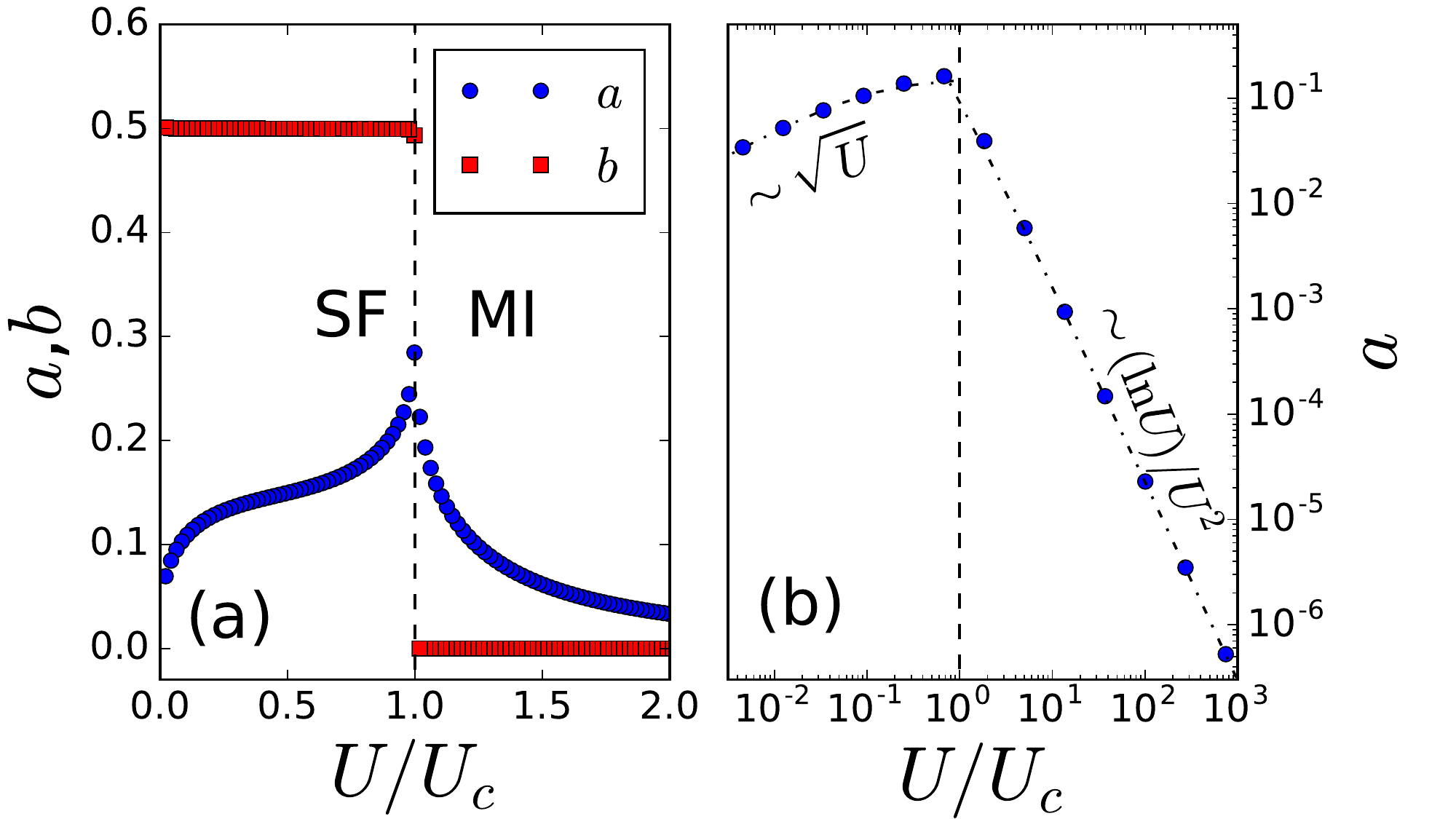}
	\includegraphics[width=1\columnwidth]{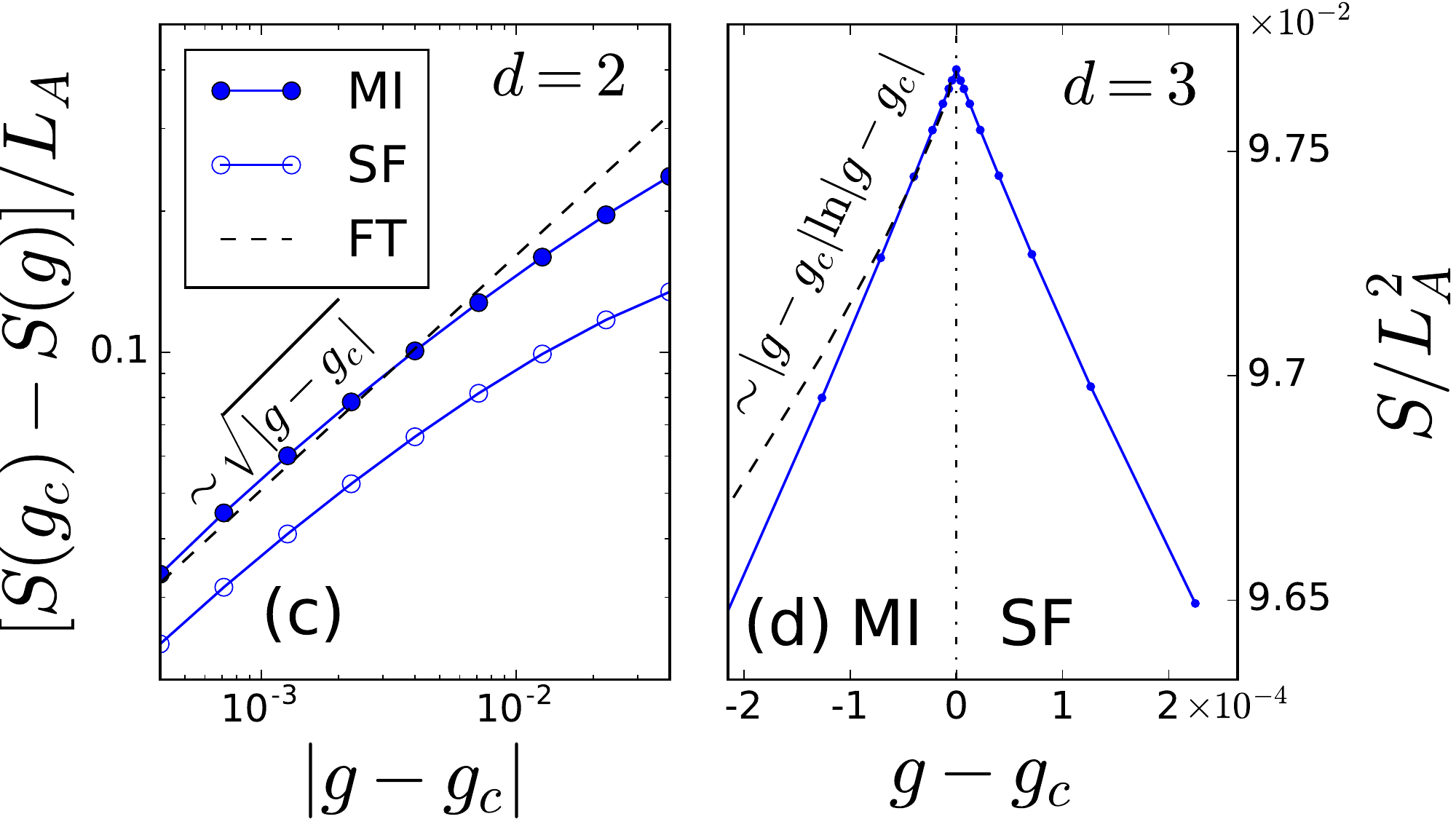}
	\caption{(a) $a$ and $b$ coefficients of EE scaling across the 2$d$ SF-MI transition at $n=1$, extracted from a fit on $L_A \times 2L_A$ half toruses with $10\le L_A\le 100$; (b) Area-law coefficient $a$ on a log-log scale. Shown are the the predictions of Bogoliubov theory for small $U$, and the large-$U$ prediction of Ref.~\cite{Alba2013} (dash-dotted line); (c) Singular behavior of the EE around the transition in $d=2$, with $L_A=100$; (d) Same as in (c) but for $d=3$ and region $A$ given by half of a $40\times 20 \times 20$ hyper-torus. For both (c) and (d), dashed line is the prediction of field theory (FT) ~\cite{Metlitskietal2009,CalabreseC2004}  (see ~\cite{SuppMat}).}
	\label{f.ab}
		\end{figure}
     
  \emph{Scaling of the entanglement entropy across the transition.} A more detailed analysis of the entanglement entropy shows that it always obeys an area-law scaling, including at the critical points. In particular, it can be fitted to the following form: 
  \begin{equation}
  S_A = a L_A^{d-1} + b \ln L_A + c 
  \end{equation}
  The $J/U$-dependence of the area-law coefficient $a$  and of the logarithm coefficient $b$ are predicted in a robust manner by the SB approach \cite{SuppMat}, and plotted in Fig.~\ref{f.ab}(a). In particular the coefficient $a$ exhibits the cusp singularity associated with the gapless Higgs entanglement mode; this is perfectly consistent with the localized nature of this mode along the boundary (see \cite{SuppMat}), making it an effective $d-1$ dimensional mode whose contribution to the entanglement entropy can only be extensive in the boundary size, $2L_A^{d-1}$.  Our approach allows us to predict as well the behavior of the area-law coefficient away from the transition. As shown on Fig.~\ref{f.ab}(b), $a\sim \sqrt{U}$ for small $U$, consistent with the prediction of Bogoliubov theory \cite{FrerotR2015} and with the fact that the area law term disappears in a perfect condensate; for large $U$ deep in the insulating phase the coefficient instead decreases as $\ln U/U^2$, in complete agreement with recent analytical and numerical calculations \cite{Alba2013} -- see also the \cite{SuppMat} for the calculation of $a$ at non-integer filling. 
 On the other hand, the coefficient $b$ of the logarithmic term takes the universal value $(d-1)/2$ throughout the superfluid phase, and jumps to zero at the phase transition. This is perfectly consistent with the prediction of Ref.~\cite{Metlitski-Grover} of a universal logarithmic term $N_G(d-1)/2 \ln L_A$ in a phase breaking a continuous symmetry, and stemming from the number of Goldstone modes $N_G$ (=1 in the case at hand). 
 
 Traditionally the CI and O(2) transitions are distinguished via the different critical exponents manifested by the order parameter, its susceptibility and the correlation length \cite{Fisheretal1989}. In fact, the  entanglement entropy is also sensitive to the critical behavior of the correlation length: indeed the area-law coefficient at the O($N$) transition is predicted \cite{Metlitskietal2009, CalabreseC2004} to manifest the singular behavior {(up to logarithmic corrections in $d=3$, see \cite{SuppMat})}
 \begin{equation}
 a = a_0(g) - a_1 \xi^{-(d-1)} =  a_0(g) - a'_1 |g-g_c|^{\nu(d-1)}
\label{e.a_sing}
 \end{equation}
  where $g = Jz/U$ is the parameter driving the O(2) transition, $a_0(g)$ is a smooth function of $g$, $\nu$ is the critical exponent of the correlation length $\xi$, and $a_1, a_1' > 0$. This behavior is verified by our data with $\nu=1/2$ -- see Fig.~\ref{f.ab}(c-d) -- as well as by previous studies \cite{Singhetal2012,HelmesW2014}. For O($N$) models with $N>1$, Eq.~\eqref{e.a_sing} as obtained in Ref.~\cite{Metlitskietal2009} actually applies only to the side of the transition on which $\xi$ diverges, namely the insulating side in the $N=2$ case at hand; and, within the Gaussian approximation, the field theory predicts quantitatively the $a'_1$ coefficient of our calculations (see Fig.~\ref{f.ab}(c-d), and \cite{SuppMat} for a detailed discussion). Yet it does \emph{not} predict the entanglement cusp observed at the O(2) transition. The existence of this cusp is in fact controlled by the behavior on the superfluid side, dominated by the presence of the amplitude mode going gapless both in the physical as well as in the entanglement spectrum. Interestingly, as shown in Fig.~\ref{f.ab}(c-d) the cusp singularity obeys Eq.~\eqref{e.a_sing} also on the superfluid side of the transition. This suggests that the entanglement entropy reveals the divergence of the elusive correlation length $\xi_{||}$ of longitudinal fluctuations \cite{Kardarbook}, which is proportional to the inverse mass of the amplitude mode (see Fig.~\ref{f.EE}(a)), but which is challenging to extract from any microscopic observable of the original model.  
 
  The sensitivity of the entanglement entropy to the softening of the amplitude mode at the O(2) transition is in fact shared with bipartite fluctuations of the particle number \cite{Racheletal2012}, but with an important difference. Such fluctuations obey the scaling $\langle \delta^2 N_A \rangle =  {\cal A} L_A^{d-1} \ln L_A + {\cal B} L_A^{d-1} + \cdots$ in the superfluid phase, namely a logarithmically violated area law. At the superfluid to insulator transition ${\cal A} \to 0$, leaving out a strict area law in the insulating phase: this aspect dominates the behavior of fluctuations across the transition (see ~\cite{SuppMat} for details). At the $O(2)$ transition, nonetheless, the ${\cal B}$ coefficient may exhibit a singularity, in analogy with what has been observed numerically at the O(3) transition in Ref.~\cite{Racheletal2012}, but, unlike in the case of the entanglement entropy, such a singularity intervenes in the subdominant terms of fluctuations and hence is not immediately visible.  


  \emph{Implications for the Higgs mode.}  The nature of the amplitude mode at the $(d+1)$-dimensional O($N$) transition with $N>1$ has been the subject of several recent studies. The mode is expected to be sharp at the transition for $d=3$, and verified to be so experimentally for $N=3$ via neutron scattering on a quantum antiferromagnet under pressure \cite{Rueggetal2008} (see also Ref.~\cite{Bissbortetal2011} for a cold-atom experiment); hence the slave boson prediction of a sharp amplitude mode is certainly reliable for $d=3$, along with the prediction of an entanglement cusp descending from the gapless amplitude mode. On the other hand, for $d=2$ the mode turns rather into a broad resonance because it decays into Goldstone modes, as observed both numerically \cite{Gazitetal2013,PolletP2012,RanconD2014} as well as experimentally within a cold-atom setup \cite{Endresetal2012}. Therefore one may question the reliability of our predictions, also in sight of the fact that an entanglement cusp is not observed in the numerical calculation of Ref.~\cite{Alba2013}. One may however argue that the latter calculation is performed on relatively small sizes, and the cusp manifests itself only when the area-law term becomes dominant for sufficiently large $L_A$. Moreover the quantum Monte Carlo calculation and scaling analysis of Ref.~\cite{HelmesW2014} for the O(3) transition in $d=2$ shows indeed such a singularity, suggesting that the same feature may be observed for $N=2$. Interestingly, no cusp exists in $d=1$ \cite{LauechliK2008,buonsanteV2006}, corroborating its relationship to the amplitude mode.  

  
   \emph{Conclusions.}  In this Letter we have shown that the entanglement entropy is a powerful probe of bosonic criticality in the Bose-Hubbard model. The implementation of the latter model via cold-atoms in optical lattices \cite{Greiner2002,Blochetal2008}, and the recent development of an interferometric scheme to measure the second R\'enyi entropy and mutual information \cite{Islametal2015}, promises that some of our predictions may actually be observed experimentally.  Our predictions are indeed robust to realistic conditions: as shown in the \cite{SuppMat}, the 2-R\'enyi mutual information at finite temperature displays a cusp singularity similar to that of ground-state entanglement. In particular the observation of such a cusp in $d=2$ would represent a new way to assess the role of the amplitude mode at the superfluid-insulator transition, fully complementary to the investigation of dynamical susceptibilities \cite{Endresetal2012}.

  \emph{Acknowledgements.} We thank A. Ran\c con and N. Laflorencie for very useful discussions. This work is supported by ANR (``ArtiQ" project).

\bibliography{SB-BH}

\newpage
\appendix


\section{Slave-boson approach}

\subsection{Mean-field solution and slave-boson transformation}

As already stated in the main text, the SB approach starts from the mean-field solution, based on the factorized ground-state Ansatz 
\begin{equation}
|\Psi_{\rm MF}\rangle = \prod_i \left [ \sum_{n} \frac{f^{(0)}_n}{\sqrt{n!}} (b_i^{\dagger})^n\right ] |0\rangle~.
\label{eq.GMF}
\end{equation}
The variational minimization of $\langle \Psi_{\rm MF} | {\cal H} | \Psi_{\rm MF}\rangle$ amounts to the self-consistent minimization of the single-site Hamiltonian 
\begin{equation}
{\cal H}_{\rm MF} = -zJ (\phi b^\dagger + \phi^*b) + \frac{U}{2} (b^\dagger)^2 b^2 -\mu b^\dagger b~.
\end{equation} 
In principle, the summation over $n$ in Eq.\eqref{eq.GMF} runs from zero to infinity. However, in practice, truncating the summation to some value $n_{\rm max}$ much larger than the mean occupation does not affect the results. For our calculations, we chose $n_{\rm max} = 5$.
The MF solution produces not only a variational ground-state, but a whole basis $|\Psi_{\alpha} \rangle$ ($\alpha = 0, ..., n_{\rm max}$) for the local Hilbert space, with coefficients ${\bm f}^{(\alpha)} = \{ f_{n}^{\alpha} \}$, reconstructing the unitary matrix $W$ which diagonalizes ${\cal H}_{\rm MF}$. Note that we focus our attention on homogeneous systems, but the SB treatment is straightforwardly extended to the non-homogeneous case where the microscopic parameters $J$, $U$ and $\mu$ could vary in space. In this case, the variational parameters $f^{(0)}_n$ could have a spatial dependence.

 The SB approach builds quantum fluctuations around the MF solution by introducing $(n_{\rm max} +1)$ fictitious SB operators $\beta_{i,n}, \beta^{\dagger}_{i,n}$ with the hardcore constraint $\sum_{n} \beta^{\dagger}_{i,n} \beta_{i,n}  = 1$, reconstructing the original Bose field operator as 
 \begin{equation}
 b_i =  \sum_n \sqrt{n+1} ~\beta^{\dagger}_{i,n} \beta_{i,n+1}~.
 \end{equation} 
 This transformation is akin to considering that the Fock states of physical bosons are ``generated" from a fictitious SB vacuum $|0_{\rm SB}\rangle$ by the application of a SB creation operator, $|n\rangle = \beta^{\dagger}_n |0_{\rm SB}\rangle$ -- whence the hardcore SB constraint. 
 The transformation $W$ diagonalizing the mean-field Hamiltonian rotates the SB $\beta$ operators to new $\gamma$ operators. In terms of the rotated operators the Hamiltonian takes then the form 
  \be
	{\cal H} = -J \sum_{\langle i,j\rangle}
	\left ( {\bm \gamma}_i^\dagger \widetilde{F} {\bm \gamma}_i {\bm \gamma}_j^\dagger \widetilde{F}^\dagger {\bm \gamma}_j + {\rm h.c.} \right ) + \sum_i {\bm\gamma}_i^\dagger \widetilde{G} {\bm\gamma}_i
	\label{eq_H_BU_SB}
	\ee
   where we have introduced the symbols:
   	\bearr
		\gamma_{i,\alpha}^\dagger &=& \sum_{n=0}^{n_{\max}} \langle n | \Psi_\alpha \rangle \beta_{i,n}^\dagger\\
		{\bm{\gamma}}_i &=& (\gamma_{i,0},\dots \gamma_{i,\alpha},\dots)^T \\
		\widetilde{F}_{\alpha\beta} &=& \langle \Psi_\alpha | {b}^\dagger | \Psi_\beta \rangle \\
		\widetilde{G}_{\alpha\beta} &=& \langle \Psi_\alpha |\frac{U}{2}{n}({n}-1) - \mu  n | \Psi_\beta \rangle 
	\eearr
   Within the SB representation, the interaction term becomes quadratic, while the hopping term is quartic: to be able to proceed further, a quadratic approximation is in order, and the rotation to the $\gamma$ operators is quite convenient in this respect. Indeed if one were to treat the $\gamma$ operators as classical complex fields, one would obtain that the energy minimum is attained for $\gamma_{0} = \gamma^{*}_0 = 1$ and all other $\gamma_{\alpha}$ fields being equal to zero: the corresponding minimal energy is the mean-field energy. Hence a meaningful harmonic expansion is performed around the MF minimum, namely ``condensing" the $\gamma_0$ bosons, $\gamma_0^{(\dagger)}\approx 1$, and treating the other bosonic fields as perturbations, $\gamma^{\dagger}_{\alpha\neq 0} \gamma_{\alpha\neq 0} \ll 1$. 
   The SB constraint is partially enforced via the ``Bogoliubov shift"
\begin{equation}
\gamma_{0,i}^{(\dagger)} \approx \sqrt{1 - \sum_{\alpha\neq 0} \gamma^{\dagger}_{i,\alpha} \gamma_{i,\alpha}} \approx 1 - \frac{1}{2} \sum_{\alpha\neq 0} \gamma^{\dagger}_{i,\alpha} \gamma_{i,\alpha}+ O(\gamma^4)~.
\end{equation}
 Discarding cubic and quartic terms in the $\gamma_{\alpha\neq 0}$ operators amounts to performing semi-classical expansion around the mean-field ground state. This treatment, as usual, must be self-consistent, in the sense that the average values of the non-condensed SB must be very small  in the renormalized ground state (or a thermal state at nonzero temperature):
 \begin{equation} \varepsilon =  \sum_{\alpha>0} \langle \gamma_{i,\alpha}^\dagger \gamma_{i,\alpha}\rangle \ll 1 ~.
 \label{e.condition}
 \end{equation} 
  As a result, the field operator $b_i^\dagger$ is expressed as : 
	\bearr
	b_i^\dagger &=& \sum_{\alpha,\beta} \gamma_{i,\alpha}^\dagger {\widetilde F}_{\alpha\beta} \gamma_{i,\beta} \nonumber\\
		&=&\gamma_{i,0}^\dagger {\widetilde F}_{00}\gamma_{i,0} + \sum_{\alpha>0}(\gamma_{i,0}^\dagger {\widetilde F}_{0\alpha} \gamma_{i,\alpha}+\gamma_{i,\alpha}^\dagger {\widetilde F}_{\alpha 0} \gamma_{i,0}) \nonumber\\
		&&~+~   \sum_{\alpha,\beta>0} \gamma_{i,\alpha}^\dagger {\widetilde F}_{\alpha\beta} \gamma_{i,\beta} \nonumber\\
		&\approx&{\widetilde F}_{00} \left ( 1-\sum_{\alpha >0} \gamma_{i,\alpha}^\dagger \gamma_{i,\alpha} \right )+ \sum_{\alpha>0}({\widetilde F}_{0\alpha} \gamma_{i,\alpha} + \gamma_{i,\alpha}^\dagger {\widetilde F}_{\alpha 0})\nonumber\\
		&&~ +  \sum_{\alpha,\beta>0} \gamma_{i,\alpha}^\dagger {\widetilde F}_{\alpha\beta} \gamma_{i,\beta} + O(\gamma^3)~.
	\eearr
	In the MI regime, $\tilde{F}_{00}=0$, and one can check that $\langle b \rangle$ remains zero at the level of the SB approximation. In the SF regime, ${\widetilde F}_{00}=\phi$ is the mean-field order parameter, which is weakly renormalized by Gaussian fluctuations as long as the fundamental assumption of the SB approach, $ \gamma_{i,\alpha}^\dagger \gamma_{i,\alpha}  \ll 1$, is verified. This implies that the MI-SF phase boundaries are not modified by the Gaussian fluctuations with respect to the mean-field solution.	
	
	\subsection{Quadratic Hamiltonian and its diagonalization}

	Injecting the above expression for the field operator in the Bose-Hubbard Hamiltonian and keeping only terms up to quadratic order in the $\gamma$ operators, one finds the following expression : 
	\be
	{\cal H} = {\cal H}^{(0)}+{\cal H}^{(1)} + {\cal H}^{(2)} + O(\gamma^3)
	\ee
	where :
	\bearr
		{\cal H}^{(0)} &=& -NJz |\widetilde F_{00} |^2 + N \widetilde G_{00}\\
		{\cal H}^{(1)} &=& \sum_i \sum_{\alpha>0} \gamma_{i,\alpha} \langle \Psi_0 | {\cal H}_{\rm MF}|\Psi_\alpha\rangle + {\rm h.c.}\\
		{\cal H}^{(2)} &=& {\cal H}_{site} + {\cal H}_{hopping} + {\cal H}_{pairs}~.
	\eearr
	${\cal H}^{(0)}$ is an overall energy offset, while ${\cal H}^{(1)}$ vanishes by definition of the $|\Psi_\alpha\rangle$ (the latter represents a useful check of the numerical calculation). ${\cal H}^{(2)}$ is composed of the three terms:
	\bearr
		{\cal  H}_{site} &=& \sum_i \sum_{\alpha,\beta>0} \gamma_{i,\alpha}^\dagger A^{(0)}_{\alpha \beta}\gamma_{i,\beta} \\
		{\cal H}_{hopping} &=& \sum_{\langle i,j\rangle} \sum_{\alpha,\beta>0}  \gamma_{i,\alpha}^\dagger A^{(1)}_{\alpha \beta}\gamma_{j,\beta}\\
		{\cal H}_{pairs} &=&  \frac{1}{2} \sum_{\langle i,j\rangle} \sum_{\alpha,\beta>0}  \gamma_{i,\alpha} B_{\alpha \beta}\gamma_{j,\beta} +{\rm h.c.}
	\eearr
	with 
	\bearr
		A^{(0)}_{\alpha\beta} &=& -\delta_{\alpha\beta}\langle \Psi_0 |{\cal H}_{\rm MF}|\Psi_0\rangle + \langle \Psi_\alpha |{\cal H}_{\rm MF}|\Psi_\beta \rangle \nonumber\\
		&=& \delta_{\alpha\beta}(\epsilon_\alpha - \epsilon_0) \label{expr_A0}\\
		A^{(1)}_{\alpha\beta}&=& -J(\widetilde F_{\alpha 0} \widetilde F_{\beta 0}^* + \widetilde F_{0\beta} \widetilde F_{0\alpha}^*) \label{expr_A1}\\
		B_{\alpha \beta} &=& -2J\widetilde F_{0\alpha} \widetilde F_{\beta 0}^* \label{expr_B}~;
	\eearr
$\epsilon_\alpha$ denote the eigenvalues of the mean-field Hamiltonian ${\cal H}_{\rm MF}$.
Since $\gamma_{i,\alpha}$ and $\gamma_{j,\beta}$ commute, we may keep only the symmetric part of $B$ : $B_{\alpha \beta} = -J(\widetilde F_{0\alpha} \widetilde F_{\beta 0}^*+ \widetilde F_{0\beta} \widetilde F_{\alpha 0}^*)$.
	We then introduce the vector notation  ${\bm\gamma}_i = (\gamma_{i,1},\dots \gamma_{i,n},\dots)^T$ (note that the operator $\gamma_{i,0}$ is not present any more), and move to Fourier space : ${\bm\gamma}_i = (1/\sqrt N) \sum_{\bm k} e^{i\bm k \cdot {\bm r}_i} {\bm\gamma}_{\bm k}$. The quadratic Hamiltonian in Fourier space takes then the form:
	\be
		{\cal H}_{2}= \frac{1}{2} \sum_{\bm k} \begin{pmatrix} {\bm\gamma}_{\bm k}^\dagger & {\bm\gamma}_{-\bm k} \end{pmatrix}
			\begin{pmatrix} A_{\bm k} & B_{\bm k}\\B_{\bm k}^* & A_{\bm k}^* \end{pmatrix}
			\begin{pmatrix} {\bm\gamma}_{\bm k} \\ {\bm\gamma}_{-\bm k}^\dagger \end{pmatrix}  + {\rm const.}
	\label{eq_H_quadra_SuppMat}
	\ee
	where we introduced the following symbols:
	\bearr
		A_{\bm k} &=& A^{(0)} + z\eta_{\bm k} A^{(1)}\\
		B_{\bm k} &=& z\eta_{\bm k} B
	\eearr
	containing the form factor $\eta_{\bm k} = (1/z) \sum_{\bm \delta} e^{i\bm k \bm \delta}$, in which the sum runs over the $z$ links originating from each lattice site. For a $d-$dimensional hypercubic lattice  
	$\eta_{\bm k} = (1/d) \sum_{i=1}^d \cos(k_ia)$ with $a$ the lattice spacing.

	The last step is to diagonalize the matrices	
	 $${\cal M}_{\bm k} = \begin{pmatrix} \mathbbm{1} &  0  \\ 0 &- \mathbbm{1}\end{pmatrix}	\begin{pmatrix} A_{\bm k} & B_{\bm k}\\B_{\bm k}^* & A_{\bm k}^* \end{pmatrix}~,$$ 
	something which is generally performed numerically given the potentially large number of SB flavors retained in the calculation.
	The Bogoliubov transformation 
	\begin{equation}
	\begin{pmatrix} {\bm{\gamma}}_{\bm k} \\ {\bm{\gamma}}_{-\bm k}^\dagger \end{pmatrix} = P_{\bm k} \begin{pmatrix} \bm{\lambda}_{\bm k}  \\ \bm{\lambda}_{-\bm k}^\dagger \end{pmatrix}
	\end{equation}
	 preserving the bosonic commutation relations \cite{blaizot}, leads finally to the diagonal form
	\be
		{\cal H}_2 = \sum_{\bm k} \sum_{\alpha=1}^{n_{\max}} \omega_{\bm k,\alpha} \lambda_{\bm k,\alpha}^\dagger \lambda_{\bm k,\alpha}~.
	\ee	
	The two lowest excitation branches of the spectrum are drawn in characteristic regions of the phase diagram in Fig.~1 of the main text.
	
		\subsection{Spectrum of the MI phase}
	
	In the MI phase, the problem  of diagonalizing the quadratic SB Hamiltonian simplifies considerably, and it lends itself to an analytical solution. Indeed, given that $\phi=0$ the mean-field Hamiltonian reduces to ${\cal H}_{\rm MF} = \sum_n \epsilon_n |n\rangle \langle n |$ with $\epsilon_n = Un(n-1)/2 - \mu n$, and the ground state is simply $|n_0\rangle$ with $n_0$ the integer minimizing $\epsilon_n$. Furthermore, $A^{(1)}$ in Eq.~\eqref{expr_A1} and $B$ in Eq.~\eqref{expr_B} only couple two SB flavors with $n=n_0\pm1$. The other modes with $|n-n_0|\ge2$ decouple, namely they diagonalize the ${\cal M}_{\bm k}$ matrix with associated eigen-energies $\omega_{\bm k,n}=\epsilon_n-\epsilon_0$ which form flat bands. For the two remaining modes with $|n-n_0|=1$, the diagonalization of the corresponding sub-matrix can be performed analytically. The corresponding eigen-energies $E_{\bm k}^{\pm}$ are the square roots of the eigenvalues of the $2\times 2$ matrix $(A_{\bm k}-B_{\bm k})(A_{\bm k}+B_{\bm k})$. The final result reads: 
	\be
		E_{\bm k}^{\pm} = \pm (\epsilon_{\bm k}/2-\delta \mu) + \frac{1}{2} \sqrt{\epsilon_{\bm k}^2 +4\epsilon_{\bm k} U x + U^2}
		\label{e.MIdispersion}
	\ee
	where $\delta \mu = \mu - U(n_0-1/2)$, $x=n_0+1/2$ and $\epsilon_{\bm k} = - Jz \eta_{\bm k}$. 
	Interestingly, this very same result can be obtained with a variety of alternative approaches, such as the random phase approximation of Ref.~\cite{SenguptaD2005}, of the time-dependent mean-field theory of Ref.~\cite{KrutitskyN2011}, all relying on quadratic expansions around the mean-field solution. Compared to previous approaches, the SB approach has the advantage of being a direct Hamiltonian approach, reducing the strongly interacting Bose-Hubbard Hamiltonian to a quadratic form, and hence lending itself most naturally to the calculation of entanglement properties.

\section{Zero mode in the SF spectrum, and its contribution to entanglement}
As already stated in the main text and in the previous section, the SB approach relies on the assumption of weak quantum fluctuations around the MF solution, corresponding to a small population of non-condensed SB on each site, Eq.~\eqref{e.condition}. Nonetheless the spectrum in the SF phase possesses a gapless Goldstone mode with linearly vanishing energy ($E_k \propto k$), giving rise to a $1/k$ divergence of the SB population. The latter could be naively interpreted as a breakdown of the theory. Yet, in the thermodynamic limit the integral $\int k^{d-1}dk/k$ over the Brillouin zone is divergent (logarithmically) only in $d=1$, while the MF solution is stable to Gaussian fluctuations in $d\ge 2$, as it is well known in the literature. As a consequence, the divergent SB population is a finite-size artefact in $d\geq 2$, which can be circumvented via different strategies.
 
The first strategy consists of simply ignoring the ${\bm k}=0$ mode in the calculations, which amounts to restricting integrals over the Brillouin zone to momenta $k\gtrsim 1/L$ where $L$ is the linear size of the system. The above procedure leads to finite-size estimates that reproduce correctly the \emph{leading} size dependence of all observables, since the discarded part in the integral is in fact of order $1/L^{d-1}$. When the zero mode is discarded, the SB approximation is very well controlled, as the SB population $\varepsilon$ remains of order $10^{-2}$  throughout the phase diagram, while raising to a maximum  of $0.11$ at the O(2) point with $n=1$ in $d=2$, and $0.04$ at the same critical point in $d=3$. 
Discarding the zero mode has no tangible effect on the entanglement spectrum of a subsystem, except for the low-energy entanglement modes; the latter ones turn out to contribute in an essential manner to the subleading scaling terms of entanglement entropy (EE) \cite{Songetal2011,Metlitski-Grover,FrerotR2015}, and hence they require a more careful treatment. 

 To correctly capture the contribution of the zero mode to the EE, one can adopt the prescription of Ref.~\cite{Songetal2011}, which consists of adding a small source term $-h(b+b^\dagger)$ to the Hamiltonian. As a result, the ${\bm k}=0$ lowest-energy mode acquires a small gap, of order $\sqrt{h}$, as it can be verified analytically within the spin-wave approximation in the hardcore-boson limit, within the Bogoliubov approximation in the weakly interacting SF regime, and numerically within the SB approach. In order to recover a gapless phase in the large $L$ limit, $h$ has to scale to zero when increasing the system size. The ${\bm k}=0$  SB hence acquire a population of order $1/\sqrt{h}$ which, redistributed over the whole system, contributes a term $O(L^{-d}/\sqrt{h})$ to the SB population $\varepsilon$ on each site. In order to preserve the condition $\varepsilon \ll 1$ the source field $h$ should therefore go to zero no faster than $L^{-2d}$. The prescription of Ref.~\cite{Songetal2011} is actually to introduce a source field scaling precisely as $h = h_0/L^{-2d}$ within the spin-wave Hamiltonian, with a prefactor $h_0$ tuned in such a way as to lead to the vanishing of the order parameter. Evidently this prescription opposes the basic assumption of spin-wave theory (namely that the classical order parameter should be weakly renormalized by quantum fluctuations); nonetheless it surprisingly enables one to recover, beyond the area law of EE, the universal subdominant contribution $N_G(d-1)/2\log L$ counting the number of Goldstone modes $N_G$ ~\cite{Metlitski-Grover}, as well as further subsubdominant geometric terms predicted theoretically ~\cite{Metlitski-Grover,Luitzetal2015,Laflorencieetal2015}. The same prescription, when applied to the SB approach, leads to an uncontrolled proliferation of the SB population, namely to the unacceptable result that $\varepsilon \gtrsim 1$. Nonetheless we observe that one can recover the universal logarithmic term $N_G(d-1)/2\log L$ within the SB approach by introducing a field $h$ scaling as $L^{-2d}$, while parametrically reducing the $h_0$ prefactor of the scaling in order to preserve the condition $\varepsilon \ll 1$. In doing this we exploit the apparent insensitivity of both area-law and logarithmic terms in the EE to the precise value of $h_0$. In particular the robustness of the logarithmic term with respect to the gap introduced in the spectrum had already been noticed by us in the context of spin-wave theory \cite{FrerotR2015}. Hence, within both the spin-wave as well as the SB approach, a fully self-consistent derivation of the two leading terms in the EE, respecting the main assumptions of the corresponding approximations, can be achieved. On the other hand, further subsubdominant terms in the scaling of the EE are found to depend on $h_0$, and they cannot be reliably predicted by the theory. Hence we abstain from investigating them in details.


\section{Entanglement from slave bosons}

\subsection{Entanglement Hamiltonian}
	In this section we provide further details about the calculation of entanglement properties within the SB approach. We have explained how to reduce the Bose-Hubbard Hamiltonian to an effective quadratic model describing the quantum fluctuations around the mean-field as Bogoliubov quasiparticles; the method presented hereafter is common to any quadratic model of bosons. 
		
	The crucial aspect which gives access to entanglement properties of quadratic models is represented by Wick's theorem \cite{Peschel2003}: as the Hamiltonian ~\eqref{eq_H_quadra_SuppMat} is quadratic in the $\gamma$'s operators, the reduced density matrix $\rho_A$ can be reconstructed from the knowledge of the two point correlators only. More precisely, being semi-positive definite $\rho_A$ can always be written $\rho_A = \exp(-{\cal H}_A)$ (${\cal H}_A$ is the so-called entanglement Hamiltonian). But as all correlation functions between degrees of freedom in $A$ factorize according to the prescriptions of Wick's theorem, $\rho_A$ itself is a Gaussian density matrix. ${\cal H}_A$ is thus a quadratic Hamiltonian which can be diagonalized by a Bogoliubov transformation ${U}_A$. The eigenvalues of the one body Hamiltonian associated to ${\cal H}_A$ (the entanglement spectrum $\omega_\alpha^{(A)}$) enable one to calculate the EE, and the associated eigenmodes give access to the spatial structure of entanglement between $A$ and $B$. Both the entanglement spectrum and the entanglement eigenmodes can be directly inferred from the one-body correlation functions $C_{ab} = \langle \gamma_a^\dagger \gamma_b \rangle$ and $F_{ab} = \langle \gamma_a\gamma_b \rangle$ (where indices $a$ and $b$ contain in fact both the position $i$ of the site and the SB flavor index $\alpha$) thanks to the relation ~\cite{FrerotR2015}
	\be
		 \begin{pmatrix} -\mathbbm{1} - C^* &  F\\ - F^*& C \end{pmatrix} = {U}_A \begin{pmatrix} {\rm diag}(-1- f_k) & 0 & \\ 
								0 &~ {\rm diag}(f_k)& \end{pmatrix} {U}_A^{-1}
		\label{relation_C_H}
	\ee
	where $f_k = 1/[\exp(\omega^{(A)}_k)-1]$ is the occupation of the $k$th entanglement mode.
	
	One can use the symmetries of the problem at hand to simplify further the diagonalization of the correlation matrix. First, if $A$ is half of a torus, then the total system shows a translational invariance in all the directions, while $\rho_A$ inherits this invariance in the directions parallel to the cut. In particular, there is no correlations between degrees of freedom living in different ${\bm k_{||}}$ sectors, so that one can perform the diagonalization in each ${\bm k_{||}}$ sector successively, in which the problem is effectively one-dimensional. Secondly, some further simplifications are in order when the correlation matrix is real (which is the case here), see Ref.~\cite{BoteroR2004} for details. Note finally that in the MI phase, one needs only to keep two SB modes corresponding to particle and hole excitations with respect to the mean-field, since the other modes are not hybridized by the Bogoliubov rotation, and they do not contribute to entanglement.
	
\subsection{Entanglement entropy} 
As $\rho_A$ takes a thermal form for an effective model of noninteracting bosonic quasiparticles, the von Neumann EE is simply 
	\be
		S_{vN} = \sum_k [(1+f_k)\ln(1+f_k) - f_k \ln f_k] = \sum_k S_{vN}^{(k)}
	\ee
	where $k$ is the mode index. 
On the other hand, for a generic density-matrix $\rho$, the R\'enyi entropy of order $\alpha$ is defined as :
	\be
		S_\alpha = \frac{1}{1-\alpha}\ln [\textnormal{Tr}( \rho^\alpha)]
	\ee
	Here, each mode of entanglement energy $\epsilon$ contains $n$ excitations with a probability $p_n = (1-e^{-\omega^{(A)}_k}) e^{-n\omega^{(A)}_k}$. The associated R\'enyi entropy is then : 
	\be
		 S^{(k)}_\alpha =  \frac{1}{1-\alpha}\ln \left[\frac{\left (1-e^{-\omega^{(A)}_k} \right)^\alpha}{1-e^{-\alpha \omega^{(A)}_k}} \right]
	\ee
	For $\alpha=2$ this simplifies as :
	\be
		S_2^{(k)}=- \ln[ \tanh(\omega^{(A)}_k/2)]~.
	\ee
	We have verified that both $S_2$ and $S_{vN}$ display the same features when calculated across the phase diagram of the BH model. 


\section{Entanglement eigenmodes and bulk-boundary correspondence}
\begin{figure}
	\mbox{\includegraphics[width=0.5\linewidth]{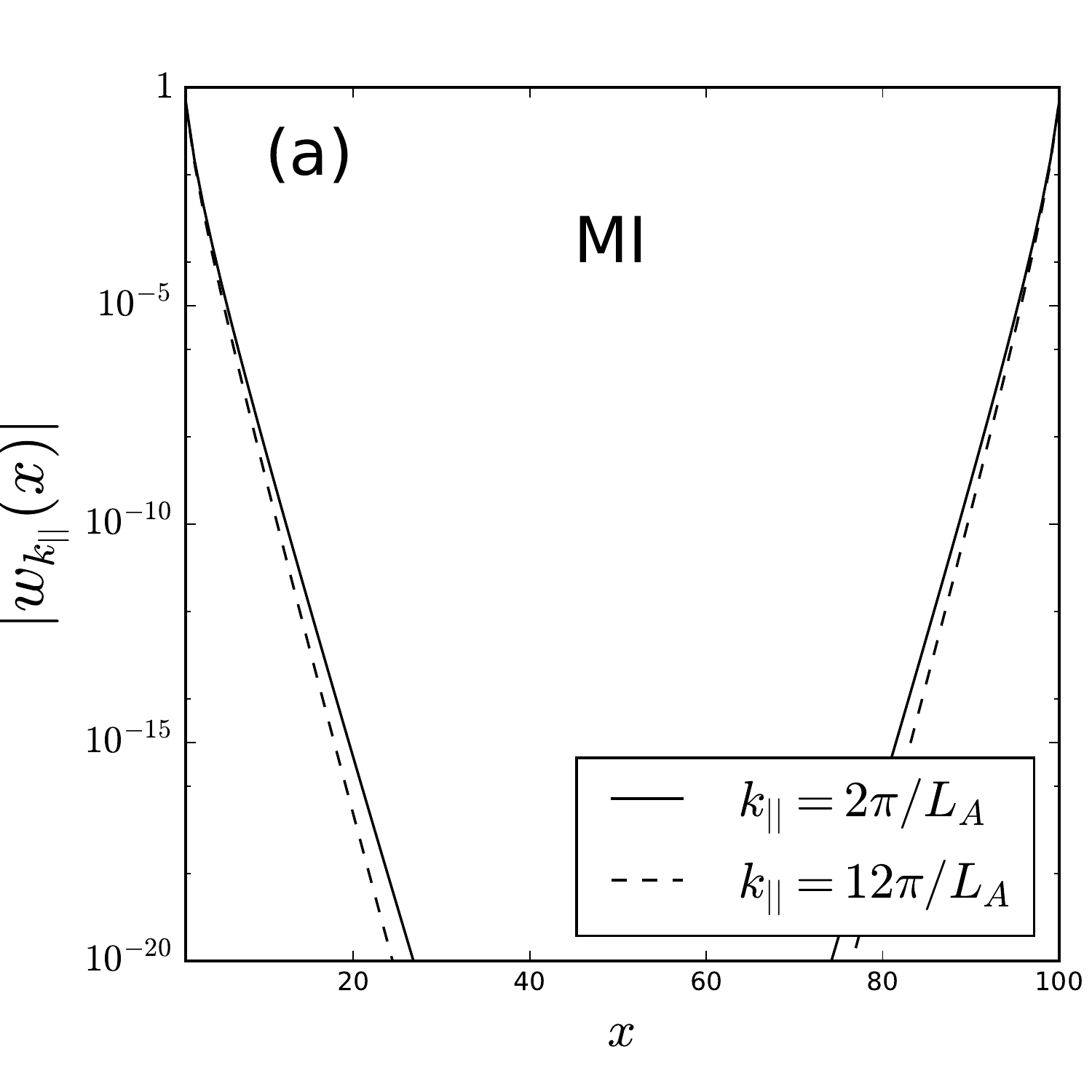} 
	\includegraphics[width=0.5\linewidth]{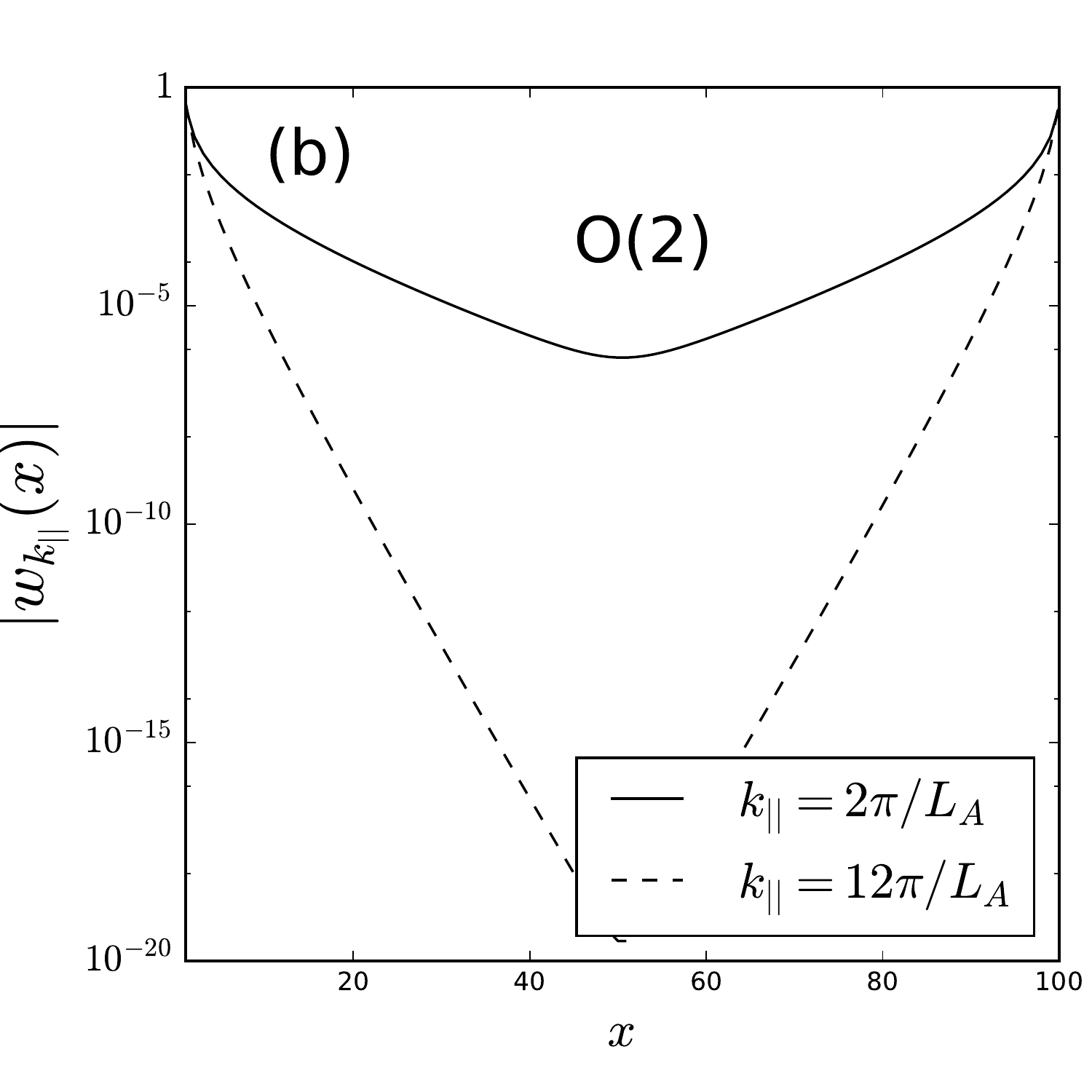} }
	\mbox{\includegraphics[width=0.5\linewidth]{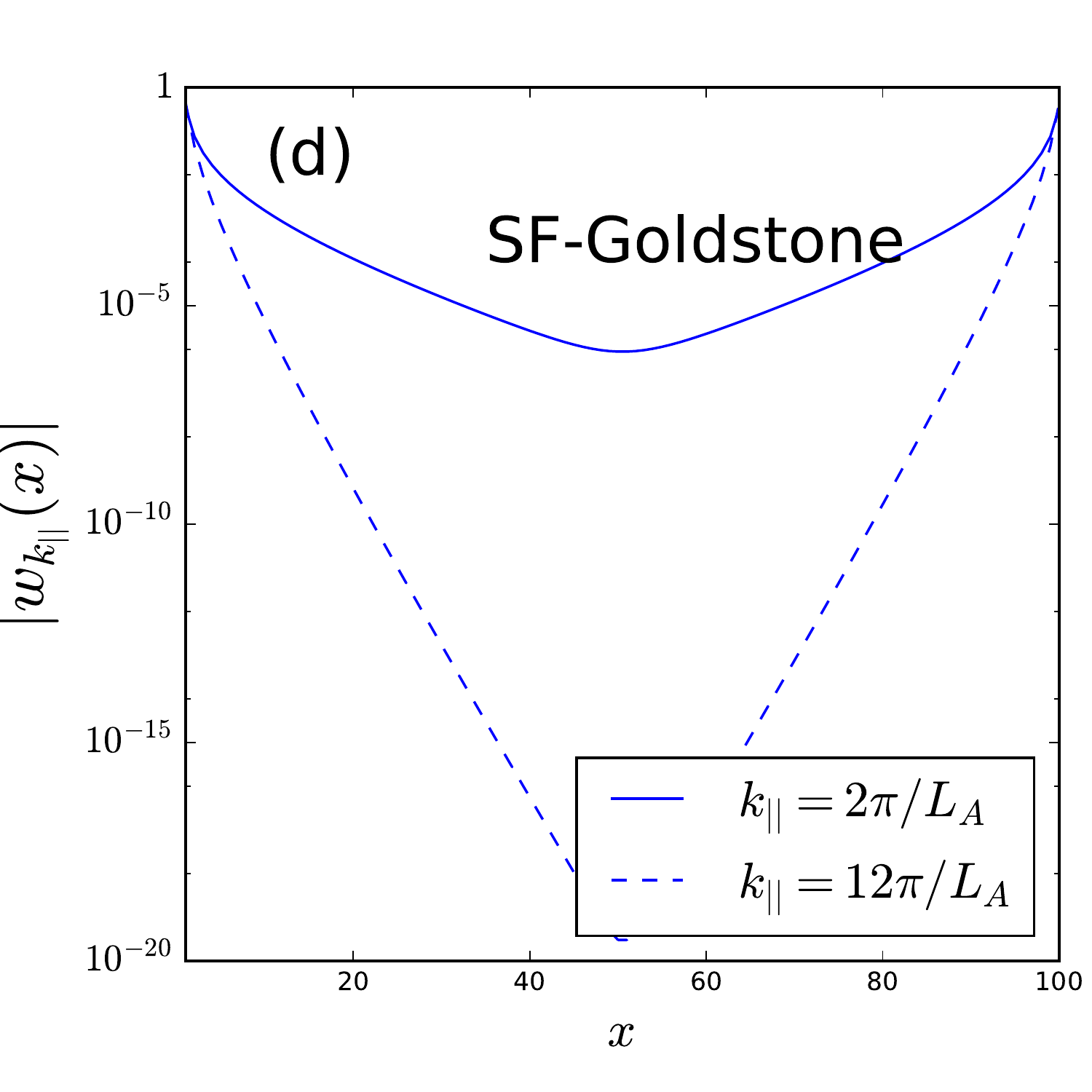} 
	\includegraphics[width=0.5\linewidth]{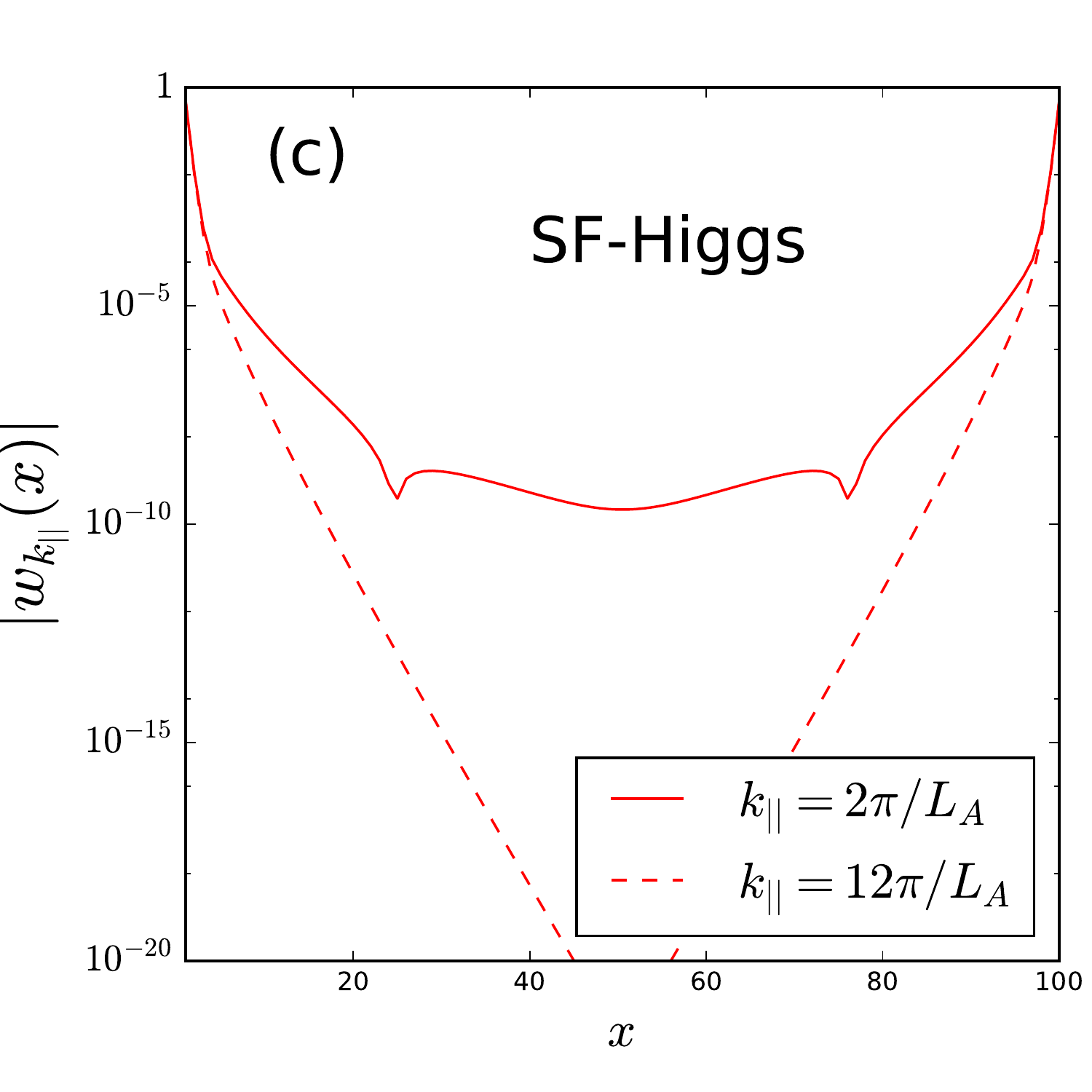}} 
	\caption{Entanglement mode profile. (a) In the MI phase; (b) at the O(2) point ; (c) in the SF phase, Higgs branch and (d) Goldstone branch. $A$ is half of a $L\times 2L$ torus with $L=100$. Parameters as in Fig.~1(a-c) in the main text. The profile is averaged over modes in the lowest corresponding branch (see text), and is plotted for ${\bm k}_{||}=2\pi/L$ and ${\bm k}_{||}=12\pi/L$.}
	\label{fig_mode_profile}
    \end{figure}

In this section, we provide further insight into the origin of the area law in the Bose-Hubbard model, by studying the spatial structure of the entanglement Higg and Goldstone-like modes. From the diagonalization of the density matrix Eq.~\eqref{relation_C_H}, one obtains the following decomposition : 
\be
	\widetilde{\lambda}_{{\bm k}_{||},\alpha}^\dagger = \sum_{x=1}^{L_A} \sum_{\beta=1}^{n_{\max}} [u_{{\bm k}_{||},\alpha}(x,\beta) \gamma_{{\bm k}_{||},x,\beta}^\dagger + v_{{\bm k}_{||},\alpha}(x,\beta) \gamma_{{\bm k}_{||},x,\beta}]
\ee
where $\widetilde{\lambda}_{{\bm k}_{||},\alpha}^\dagger$ creates a quasiparticle in the $({\bm k}_{||},\alpha)$ entanglement mode, whereas $\gamma_{{\bm k}_{||},x,\beta}^\dagger$ creates a quasiparticle at a distance $x$ from the $A-B$ boundary, with a momentum ${\bm k}_{||}$ parallel to the cut, and in the SB mode $\beta$ (remind that $A$ is periodic along all directions but $x$). The normalization condition of the $({\bm k}_{||},\alpha)$ entanglement mode imposes :
\be
	\sum_{x=1}^{L_A} \sum_{\beta=1}^{n_{\max}} [u_{{\bm k}_{||},\alpha}(x,\beta)^2 - v_{{\bm k}_{||},\alpha}(x,\beta) ^2] = 1
\ee
so that the weight of each mode at a distance $x$ from the $A$-$B$ cut is naturally introduced as : 
\be
	w_{{\bm k}_{||},\alpha} = \sum_{\beta=1}^{n_{\max}} [u_{{\bm k}_{||},\alpha}(x,\beta)^2 - v_{{\bm k}_{||},\alpha}(x,\beta) ^2] ~.
\ee
This is the function plotted on Fig.~\ref{fig_mode_profile}. Note that when the mode is (quasi) degenerate, associated either with the existence of two boundaries for $A$, or when, in the MI phase, the Higg and Goldstone-like branches combine into a doubly degenerate particle-hole-like branch, the function plotted is actually the average of the weights $w_{{\bm k}_{||},\alpha}$ over the (quasi) degenerate corresponding modes $\alpha$.
    One sees that the modes in the lowest branch of the entanglement spectrum are exponentially localized in the vicinity of the boundary of $A$, giving a complementary insight into the physical origin of the area law.
      \footnote{The precise way in which the weight $f_{{\bm k}_{||},\alpha}$ decays when moving away from the boundary depends on the mode. In particular the modes in the ${\bm k} = 0$ sector behave very differently -- some of them displaying even a flat profile. Nonetheless they give a vanishingly small contribution to the area law in the thermodynamic limit, which is the main purpose of our study.} 
 Interestingly, the modes in the $k$-th lowest branch of the ES are found to be localized in the vicinity of the $k$-th site away from the boundary. This suggests that the entanglement Hamiltonian in $d$ dimensions has the structure of a collection of Hamiltonians in $d-1$ dimensions describing the couplings parallel to the $A$-$B$ cut at variable distance from the cut itself. These Hamiltonians have $(d-1)$-dimensional eigenstates which are mixed by coupling terms perpendicular to the cut in a highly \emph{non-resonant} manner, so that the resulting eigenstates of the full entanglement Hamiltonian bear a similar spatial structure to those of the $(d-1)$-dimensional Hamiltonians. Hence the entanglement eigenmodes can be regarded as eigenmodes of boundary Hamiltonians effectively living in $d-1$ dimensions. Within the SB approximation, the low-energy spectrum of these boundary Hamiltonians show substantial analogies with the low-energy spectrum of the original microscopic Hamiltonian, displaying the same proportions of gapped and gapless modes. The latter ones can be then classified as entanglement Higglike and Goldstone-like modes, respectively. 


\section{Entanglement in the superfluid phase: bridging Bogoliubov and spin-wave theory}

\begin{figure}
	\includegraphics[width=1.\linewidth]{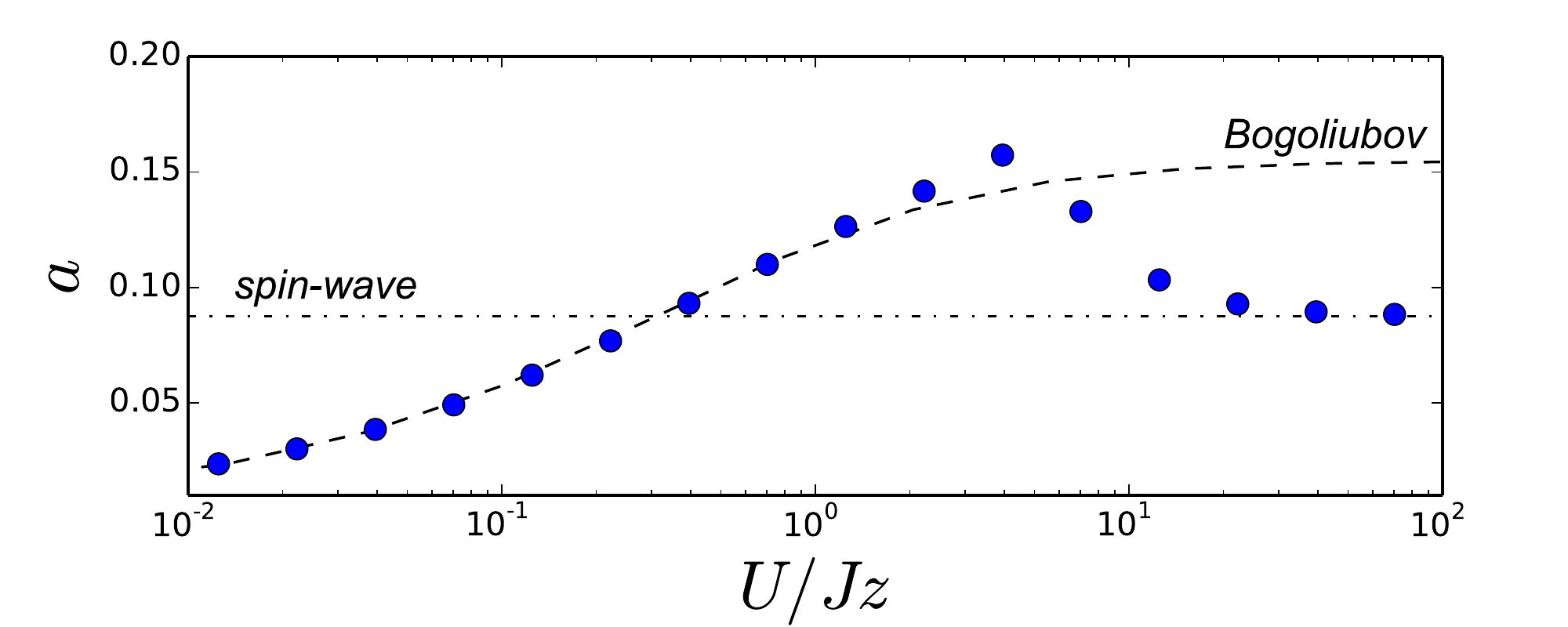}
	\caption{Entanglement entropy along the $n=0.9$ line of the phase diagram Fig.~2(c) of the main text. $a$ results from fits of the form $S=al+b\ln l+c$ for $10\le l \le 100$, and for each $l$, $A$ is half of a $l\times 2l$ torus. The dashed (dotted-dashed) line is the prediction of Bogoliubov (spin-wave) approximation.}
	\label{fig_S_n=0.9}
	\end{figure}

The EE of interacting bosons has been studied in the recent past in the opposite limits of hardcore bosons \cite{Songetal2011,Luitzetal2015,FrerotR2015} and of the weakly interacting Bogoliubov gas \cite{FrerotR2015}. Remarkably, the SB approach allows one to interpolate continuously between these extreme limiting behaviors within a single theory. Fig.~\ref{fig_S_n=0.9} shows the area-law coefficient $a$ of the EE at fixed but non-integer density $n=0.9$ over a broad range of interaction strengths $U/Jz$ (see the corresponding line on the phase diagram on Fig.~2 in the main text).
	
	In the two opposite limits $U\gg Jz$ and $U \ll Jz$, one recovers spin-wave and Bogoliubov predictions, respectively, while a naive extrapolation of the Bogoliubov and spin-wave predictions outside of their ranges of validity  does not allow to reconstruct the intermediate behavior. In general, we observe that both spin-wave and Bogoliubov approximations fail to describe the vicinity of the Mott phase: this is not surprising since they do not include the amplitude mode, which plays a major role in the vicinity of the critical point.

\section{Particle number fluctuations from slave bosons}
  \begin{figure}
	\includegraphics[width=1\linewidth]{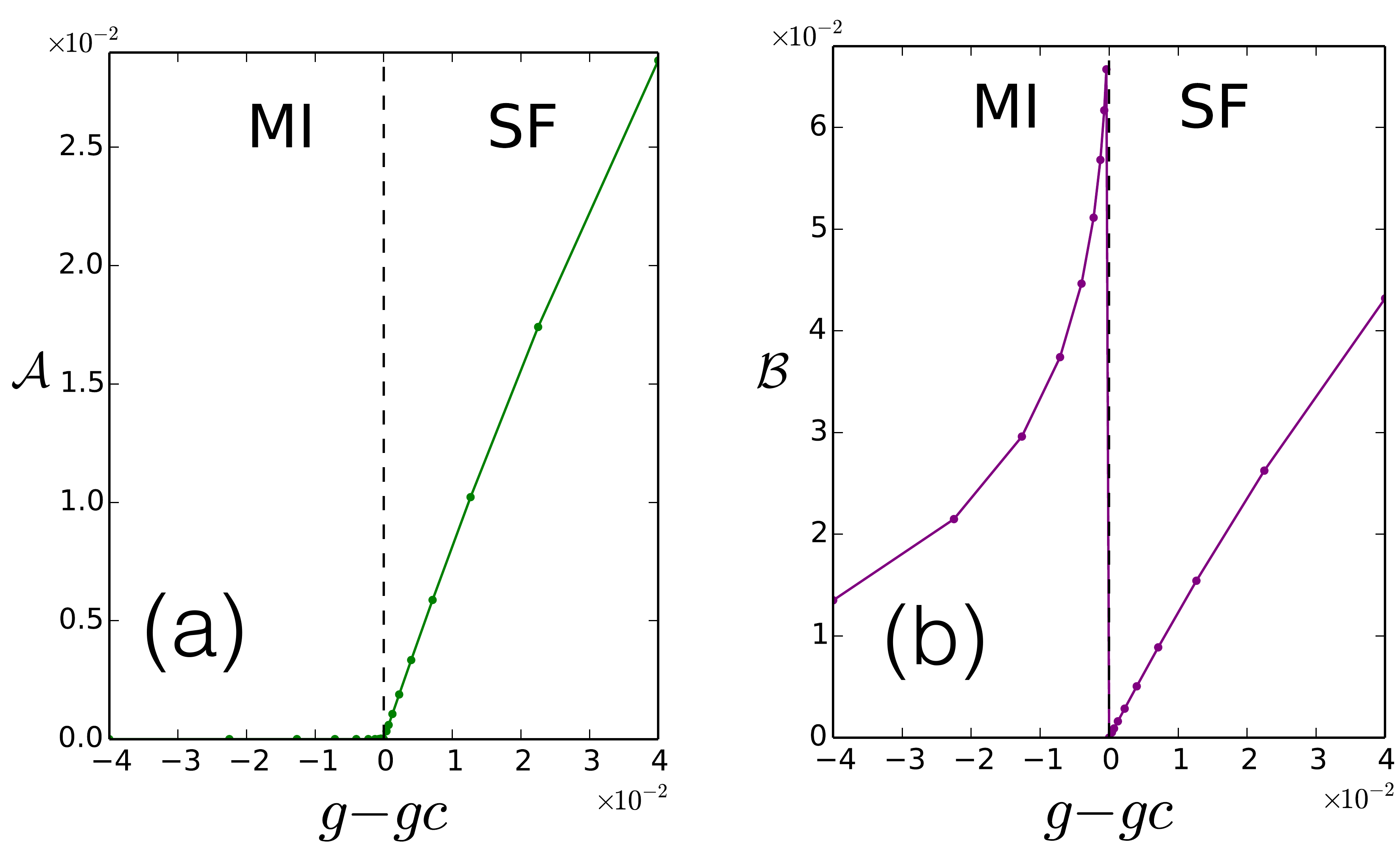}
	\includegraphics[width=1\linewidth]{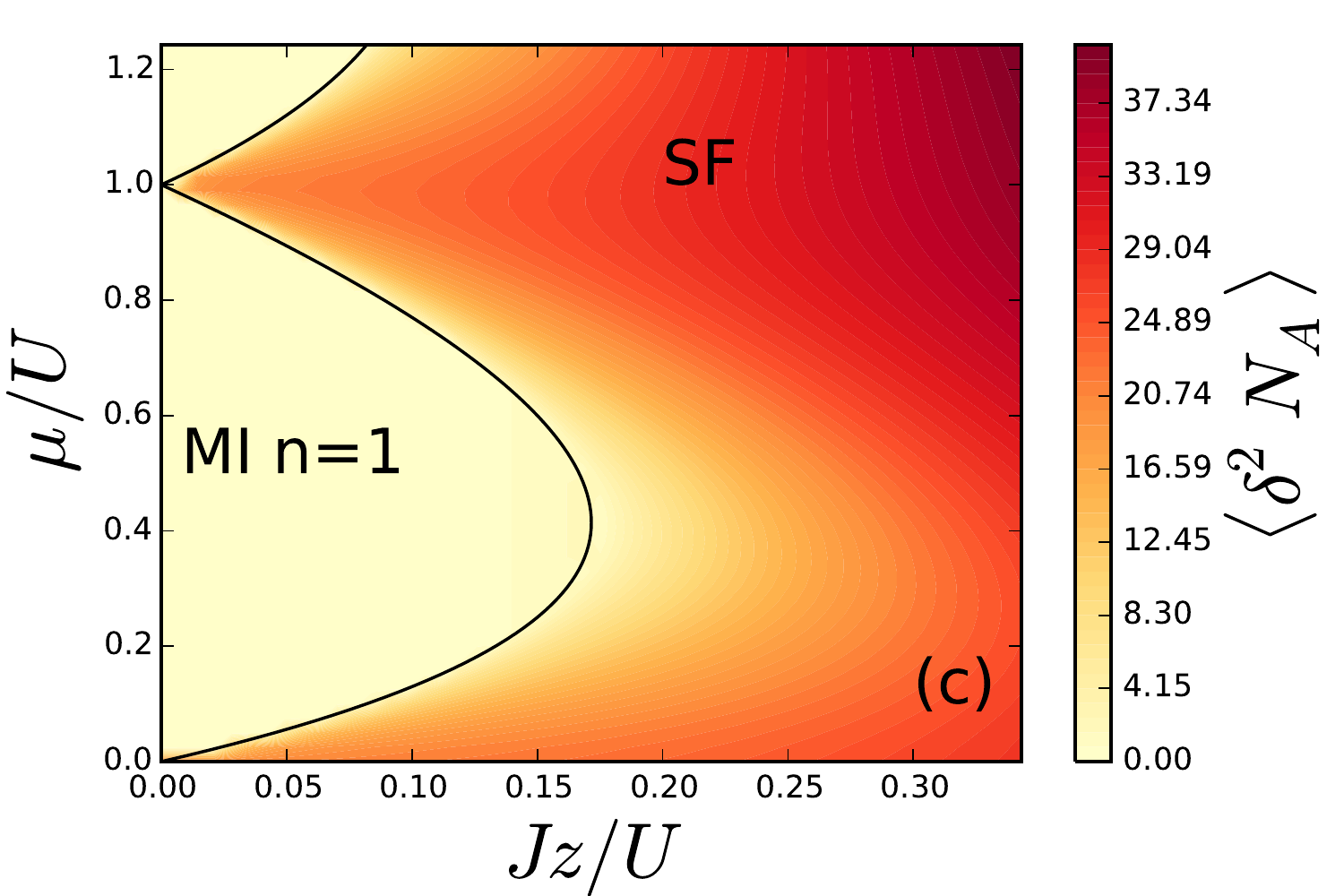}
	\caption{Bipartite fluctuations of the number of particles. (a) and (b) : ${\cal A}$ and ${\cal B}$ coefficients resulting from a fit of the form $\langle \delta^2 N_A \rangle = {\cal A} L \log L +{\cal B} L + {\cal C}$ where $A$ is half of a $L\times 2L$ torus. (c) $\delta^2 N_A$ over the phase diagram for $l=50$.}
	\label{fig_fluctuations}
	\end{figure}

	In this section we provide the details of the calculation of the density-density correlation function $\langle \delta n_i \delta n_{j} \rangle$ within the SB approach. The knowledge of density correlations allows one in turn to reconstruct the scaling of particle-number fluctuations. In the SB representation the density operator takes the form :
	\be
		n_i = \sum_n  n \beta_{i,n}^\dagger \beta_{i,n} 
	\ee
	or, in terms of the $\gamma$ operators :
	\be
		n_i =  \bm{\gamma_i}^\dagger \widetilde{N} \tilde{\bm\gamma_i}
	\ee
	with $\widetilde{N}_{\alpha \beta} = (W^\dagger N W)_{\alpha \beta} = \langle \Psi_\alpha|b^\dagger b|\Psi_\beta\rangle$. When treating density correlations, we have to distinguish between the MI phase and the SF phase. In the MI phase, the $\gamma$ operators are labeled by the on-site particle number $n$, and the density fluctuation operator takes the form:
	\be
		\delta n_i = n_i - n_0 =  \sum_{n\neq n_0} \gamma_{i,n}^\dagger (n-n_0) \gamma_{i,n} 	
	\ee	
	where $n_0$ is the filling fraction, and we used the condition $\sum_n \gamma_{i,n}^\dagger \gamma_{i,n} = 1$. Density correlations $\langle \delta n_i \delta n_{j} \rangle = \langle n_i n_j \rangle - n_0^2$ can then be calculated using Wick's theorem  
	\bearr
		\langle \delta n_i \delta n_j \rangle_{\textnormal{MI}} = \sum_{n,n'\neq n_0} (n-n_0)(n'-n_0) ~~~~~~~~~~\\ \nonumber
\left[ \langle \gamma_{i,n}^\dagger  \gamma_{j,n'}^\dagger\rangle  
\langle \gamma_{i,n}  \gamma_{j,n'}\rangle +  \langle \gamma_{i,n}^\dagger  \gamma_{j,n'} \rangle \langle \gamma_{i,n}  \gamma_{j,n'}^\dagger \rangle \right]\\~.
\label{eq_fluct_MI}
	\eearr
	On the other hand, in the SF phase, some cross terms $\gamma_{i,\alpha}^\dagger\widetilde{N}_{\alpha,\beta}\gamma_{i,\beta}$ with $\alpha\neq\beta$ are present in the expression of $n_i$. Replacing $\gamma_{i,0}$
	and $\gamma^{\dagger}_{i,0}$ with $1-\frac{1}{2}\sum_{\alpha>0}\gamma_{i,\alpha}^\dagger\gamma_{i,\alpha}$, one sees that the dominant contribution in the density correlations is 
	\bearr
		\langle \delta n_i \delta n_{j} \rangle_{\textnormal{SF}} &=&\sum_{\alpha,\beta>0} \widetilde{N}_{0\alpha}\widetilde{N}_{0\beta}\langle (\gamma_{i,\alpha} + \gamma_{i,\alpha}^\dagger)(\gamma_{j,\beta} + \gamma_{j,\beta}^\dagger)\rangle \nonumber \\
		&+& O(\gamma^4)
	\label{eq_fluct_SF}
	\eearr
	where we used the fact that $\widetilde{N}$ is a real and symmetric matrix. Terms of order $O(\gamma^4)$ have to be neglected at the level of the SB approximation, as long as $O(\gamma^2)$ terms are present. The bipartite fluctuations of the number of particles  follow by integrating twice the correlation functions in Eqs.~\eqref{eq_fluct_MI} and \eqref{eq_fluct_SF}, $\delta^2 N_A = \sum_{i,j\in A} \langle \delta n_i \delta n_{j}\rangle$.\\
	From the above discussion, one immediately notices a discontinuity in the way the correlation function is calculated between the SF and MI phases -- something which is an artefact of the SB approach. Nevertheless, as shown on Fig. ~\ref{fig_fluctuations} we are able to predict the correct dominant scaling for the fluctuations : $\delta^2 N_A= {\cal A} L_A^{d-1} \log L_A + {\cal B}  L_A^{d-1} + O(L_A^{d-2})$, with ${\cal A} =0$ in the MI phase (\textit{i.e.} a strict area law in the gapped MI phase, and a logarithmically violated area law in the critical SF phase). 
	
	In Ref.~\cite{Racheletal2012}, a singularity of the coefficient ${\cal B}$ of the subdominant area law contribution to the fluctuations has been found at the $O(3)$ quantum critical point (QCP) of the Heisenberg antiferromagnetic model on a spatially anisotropic square lattice. We would expect a similar singularity to occur at the $O(2)$ QCP of the Bose-Hubbard model, but in the SF phase, this subdominant area law contribution is unfortunately not reliably accounted for by the SB approximation. Indeed, when approaching the critical point, the ${\cal A}$ coefficient becomes smaller and smaller, so that the neglected $O(\gamma^4)$ terms in Eq.~\eqref{eq_fluct_SF} are expected to become important on finite size systems. However, including those terms in our calculation would be inconsistent with the very first step leading to the quadratic SB hamiltonian (as a matter of fact, if kept in the calculation, the quartic terms are found to give an unphysical volume contribution to the fluctuations). Nonetheless, as already discussed in the main text, the cusp singularity in the ${\cal B}$ coefficient does not manifest itself in the total fluctuations \cite{Racheletal2012}, as it is masked by the vanishing of the logarithmically violated area law which dominates the behavior of the fluctuations across the transition. This aspect is explicitly shown in Fig.~\ref{fig_fluctuations}(c), where the particle-number fluctuations across the phase diagram of the Bose-Hubbard model generically display an enhancement at the onset of the SF phase, but no cusp maximum at the SF-MI transition (be it of $O(2)$ or CI type).


\section{Field-theory prediction for the $O(N)$ model within the Gaussian approximation }
Ref.~\cite{Metlitskietal2009,CalabreseC2004} has formulated an explicit prediction for the behavior of the EE in the disordered phase of the $O(N)$ model within the Gaussian approximation. Via universality, such a prediction should also apply quantitatively to our calculation in the vicinity of the SF-MI phase approached from the MI side.  The Lagrangian of the O(N) field theory within the Gaussian approximation has the form :
\be 
	L = \frac{1}{2}\sum_\mu (\partial_\mu \phi)^2 + \frac{m^2}{2} \phi^2
\ee
where $\mu = x_1, x_2, ...., x_d, \tau$ indexes the $(d+1)$ dimensions of Euclidean space-time. 
The dispersion relation is $\omega^2 = k^2 + m^2$. This induces a correlation length $\xi = m^{-1}$, which in turn gives a singular (negative) contribution to the EE 
\begin{equation}
\frac{S_{sing}}{\cal A} = \frac{r}{\xi^{d-1}} = rm^{d-1}
\label{E.s_sing}
\end{equation}
  where $r$ is a coefficient dependent on $d$ and $N$ ($\cal A$ is the area of the boundary between $A$ and $B$). This contribution is singular because $\xi^{-1} \sim \sqrt{\vert g-g_c \vert}$ within the Gaussian model, with $g=Jz/U$ and $g_c = 3-2\sqrt{2}$ in the MF treatment of the BH model. In fact, the prediction \eqref{E.s_sing} of Ref~\cite{Metlitskietal2009} is only valid below the upper critical dimension $d_c=3$. Indeed, an explicit prediction for the coefficient $r$ is obtained from the Eq. (6.6) of Ref ~\cite{CalabreseC2004}, namely :
  \be
  S/{\cal A} = -\frac{N}{12}\int\frac{{\rm d}^{d-1}k_{||}}{(2\pi)^{d-1}} \log\frac{k_{||}^2+m^2}{k_{||}^2+a^{-2}} ~,
\ee  
 where the integration runs over the $d-1$ spatial dimensions of the boundary, and $a$ is the lattice spacing. The singular part comes from the small $m$ terms resulting from the integration, namely : 
 \be
 S_{sing}/{\cal A} = -\frac{N}{12}m ~~~~~ (d=2) 
 \ee
 \be
 S_{sing}/{\cal A} = \frac{N}{48\pi} m^2\ln m^2 ~~~~~ (d=3),
 \ee
 so that a logarithmic correction to Eq.~\eqref{E.s_sing} appears in $d=3$.
 Since the dispersion relation of the Bose-Hubbard model in the Mott phase near the O(2) point is $\omega^2 = k^2 c^2 + \Delta^2$, we can identify $m$ with $\Delta/c$. The sound velocity $c$ at the critical point and the gap $\Delta$ in the Mott phase are obtained by expanding the full dispersion relation of Eq.~\eqref{e.MIdispersion} (with $z=2d$) :
\begin{equation}
\Delta^2 = U^2\sqrt 2{\vert g_c-g\vert} ~~~~~~~~~ c^2 = U^2 (g_c/z)\sqrt 2 ~.
\end{equation}
 Hence, the Gaussian field-theory prediction for the Bose-Hubbard model would be:
\be
	S_{sing}/{\cal A} = -\frac{1}{3}\sqrt{\left| \frac{g}{g_c}-1 \right|} ~~~(d=2)
\ee
\be
	S_{sing}/{\cal A} = \frac{1}{4\pi} \left| \frac{g}{g_c}-1 \right| \ln \left( 6\left| \frac{g}{g_c}-1 \right| \right)~~~(d=3).
\ee
As seen on Fig.~3(c-d) of the main text, this prediction agrees very well with our data upon approaching the transition from the MI side.


\begin{center}
  \begin{figure}[ht!]
	\includegraphics[width=\linewidth]{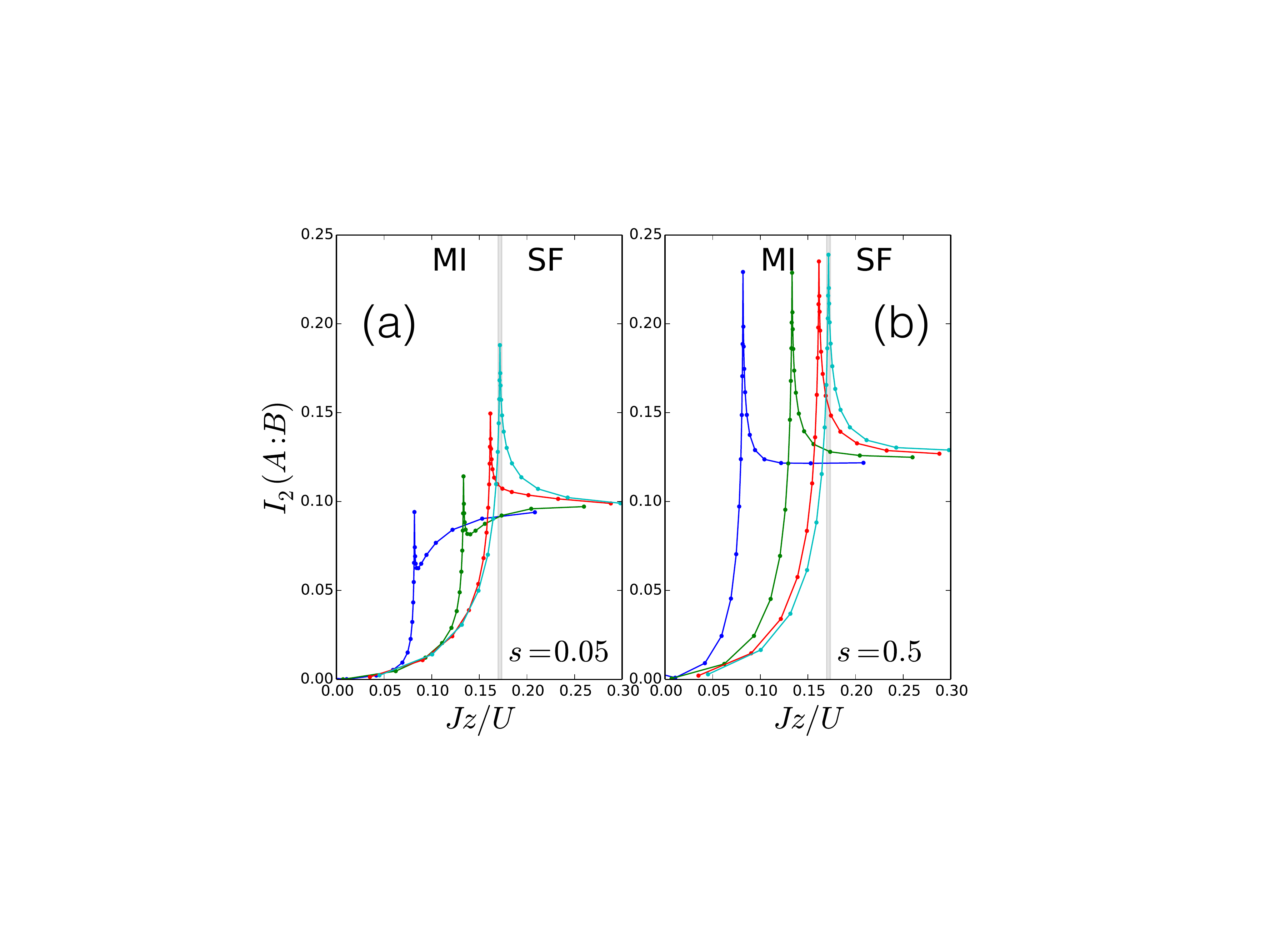}
	\caption{2-R\'enyi entropy at finite thermal entropy per particle $s$ for the 2$d$ Bose-Hubbard model across the SF-MI transition for different fixed chemical potentials, $\mu/U=0.1$, $0.2$, $0.3$ and $\sqrt 2 -1$. Shaded area same as in Fig.~2(b) of the main text.
	(a) $s=0.05$; (b) $s=0.5$. Region $A$ is half of a $30\times 60$ torus.}
	\label{fig_I2}
	\end{figure}
	\end{center}

\section{Cusps in the 2-R\'enyi mutual information at finite temperature}

In order to formulate a prediction for potential experiments in cold-atom quantum simulators, we consider the behavior of the 2-R\'enyi mutual information
$I_2(A:B) = S_2(A) + S_2(B) - S_2(A+B)$, which has been recently measured in Ref.~\cite{Islametal2015}. Cold-atom experiments are generically performed at fixed thermal (=von Neumann) entropy density $s$ -- controlled by the initial conditions of the experiments -- rather than at fixed temperature. Hence we consider the evolution of the mutual information at fixed $s$ across the MI-SF transition, as showns in Fig.~\ref{fig_I2}. We observe the remarkable feature that the cusp singularity observed in the entanglement entropy at $T=0$ or $s=0$ at the O(2) transition is preserved in the mutual information, implying that it is an observable feature under realistic conditions. We moreover observe that for $s>0$ a cusp develops also at the CI transition, attribuable to the thermal population of the amplitude mode being softest at the transition point. At sufficiently high entropy density, a cusp singularity of the mutual information becomes a characteristic of the whole critical line in the phase diagram of the Bose-Hubbard model, showing that it can serve as a very sharp probe of the SF-MI transition.

\end{document}